\documentclass[preprint]{aastex}

\usepackage{color}
\usepackage{amsmath}
\usepackage{mathrsfs}
\usepackage{stmaryrd}
\usepackage{multirow}
\usepackage{hyperref}
\usepackage{pdflscape}
\usepackage{ulem}

\def\be{\begin{equation}}
\def\ee{\end{equation}}
\def\bd{\begin{displaymath}}
\def\ed{\end{displaymath}}
\def\ba{\begin{aligned}}
\def\ea{\end{aligned}}
\def\nms{\mathsurround=0pt}
\def\oversim#1#2{\lower 4pt\vbox{\baselineskip 0pt \lineskip 1pt
    \ialign{$\nms#1\hfil##\hfil$\crcr#2\crcr\sim\crcr}}}

\def\bh{M_{\bullet}}

\def\g{{\rm g}}

\def\msun{M_{\odot}}
\def\AU{\rm ~AU}
\def\kms{\rm ~km~s^{-1}}
\def\max{{\rm max}}

\def\orb{{\rm orb}}
\def\obs{{\rm obs}}

\def\p{{\rm p}}

\def\c{{\rm c}}
\def\s{{\rm s}}
\def\rc{{\rm rc}}
\def\zc{{\rm zc}}
\begin{document}

\title{On the Newtonian and spin-induced perturbations felt by the stars orbiting around the 
massive black hole in the Galactic Center}

\author{Fupeng Zhang$^{1,\dagger}$ and Lorenzo Iorio${^2}$}

\affil{ $^1$ School of Physics and Astronomy, Sun Yat-Sen University, Guangzhou 510275, 
China;~$^\dagger$\, zhangfp7@mail.sysu.edu.cn\\
$^2$  Ministero dell' Istruzione, dell' Universit\`a e della Ricerca 
(M.I.U.R.)-Istruzione, Fellow of the Royal 
Astronomical Society (F.R.A.S.) Viale Unit\`a di Italia 68, 70125, Bari (BA), Italy; 
lorenzo.iorio@libero.it }
\begin{abstract}
The S-stars discovered in the Galactic center (GC) are expected to provide unique 
dynamical tests of the Kerr metric of the massive black hole (MBH) orbited by them. In order to 
obtain unbiased measurements of its spin and the related relativistic effects, a comprehensive 
understanding of the gravitational perturbations of the stars and stellar remnants around the MBH 
is quite essential. Here, we study the perturbations on the observables of a typical target star, 
i.e., the apparent orbital motion and the redshift, due to both the spin-induced relativistic 
effects and the Newtonian attractions of a single or a cluster of disturbing object(s). We find 
that, in most cases, the Newtonian perturbations on the observables are mainly attributed to the 
perturbed orbital period of the target star, rather than the Newtonian orbital precessions. The 
Newtonian perturbations have their unique features when they peak around the pericenter passage in 
each revolution, which is quite different from those of the spin-induced effects. Looking at the 
currently detected star S2/S0-2, we find that its spin-induced effects on both the image position 
and redshift are very likely obscured by the gravitational perturbations from the star S0-102 
alone. We also investigate and discuss the Newtonian perturbations on a hypothetical S-star located 
inside the orbits of the currently detected ones. By considering a number of possible stellar 
distributions near the central MBH, we find that the spin-induced effects on the apparent position 
and the redshift dominate over the stellar perturbations for target stars with orbital semimajor 
axis smaller than $100-400\AU$ if the MBH is maximally spinning. Our results suggest that, in 
principle, the stellar perturbations can be removed as they have distinctive morphologies comparing 
to those of the relativistic Kerr-type signatures.
\end{abstract}


\section{Introduction}
It is now widely accepted that a massive black hole (MBH) exists in the center of 
our galaxy, with the most prominent evidence provided by the so-far Keplerian motion of 
dozens of the surrounding S-stars~\citep{Ghezetal08,Gillessenetal09}. They are found to be 
exclusively B-type dwarfs, and closely orbiting the central MBH within a distance of 
$\sim0.04$pc$\simeq8250\AU$. The continuous monitoring of their orbital motion provides
precise measurements of the mass of the central MBH ($\simeq4\times10^6\msun$) and, simultaneously,  
of the Galactic distance ($\simeq8$\,kpc)~\citep{Ghezetal08,Gillessenetal09,Meyer12}. Theoretical 
studies suggest that some of the hidden S-stars exist within the orbits of the currently detected 
ones, which may be revealed by the future telescopes, e.g., the thirty meter telescope (TMT) 
or European extremely large telescope (E-ELT)~\citep[e.g.,][]{Zhang13}. Due to the 
proximity of these S-stars to the MBH, the strength of the gravitational field around 
them is orders of magnitude larger than those in the solar system and the pulsar 
binaries~\citep{Angelil10a,Iorio11a}. Thus, their trajectories contain 
various general relativistic (GR) effects, including the Lense-Thirring precession and the frame 
dragging~\citep[e.g.,][]{Jaroszynski98,Fragile00,RE01,Weinberg05,Will08,PS09,Angelil10a,AS10,AS11,
Merritt10,Iorio11a, Iorio11b,Zhang15}, which should be measured accurately by the powerful 
facilities in the near future. The continuous tracking of the orbital motion of 
the S-stars by the future telescopes is expected to provide unique dynamical tests of the 
Kerr metric of the MBH and also the no-hair theorem in the Galactic 
center (GC)~\citep[e.g.,][]{Zhang15,Psaltis15,Johannsen16,Yu16}.

However, in order to make the accurate measurements of the spin and its induced GR effects 
actually feasible, a careful handling of the perturbations induced by the other stars on the 
motion of the target ones (The so-called ``Newtonian perturbation'' or the ``stellar 
perturbation'') is required. Indeed, they are likely contributed by a number 
of different gravitational sources located in the vicinity of the 
target star in the GC, e.g., the late-type and early-type stars~\citep[e.g.,] 
[]{Paumard06,Gillessenetal09,Bartko10}, stellar mass black holes, neutron stars, white 
dwarfs~\citep[e.g.,][]{Freitag06,Morris93}, and dark matters~\citep[e.g.,][]{Iorio13}. An 
intermediate mass black hole (IMBH) possibly exists, with some allowed parameter space 
for its orbit and mass according to the 
current observations~\citep[e.g.,][]{HM2003,Yu03,Genzel10,Gualandris09,Gillessenetal09,
Gualandris10}. Other dynamical processes, e.g., gravitational wave and tidal dissipation, 
are important but only for those S-stars in highly eccentric orbits and/or in extremely 
tight orbits~\citep{Psaltis12,Psaltis13,Psaltis15}.

The stellar perturbation can induce additional orbital precessions of the target star and 
submerge those due to the spin-induced effects. N-body post-Newtonian numerical  
simulations~\citep{Merritt10} or the analytical estimations base on the orbital 
perturbing theories~\citep{Sadeghian11}, both found that the orbital precessions of the 
target star caused by the stellar perturbations can obscure those due to the 
frame-dragging effects (or the quadrupole effects) if the target star itself is located 
outside distance of $\sim0.5$mpc (or $\sim0.2$mpc) from the MBH. However, these previous studies 
have not included other complexities. For example, the precessions of the ascending node, periapsis 
and the orbital inclination can only be indirectly determined by fitting 
the predictions of models incorporating the various GR
effects and also the complexities due to the MBH parameters
(e.g., mass, spin and GC distance)  to the directly observables of the target star, 
i.e., the apparent trajectories in the plane of the sky and the redshift.
Thus, from a practical point of view, it is more meaningful
to compare the predicted perturbations on the apparent trajectories
and redshifts of the target star due to spin effects
to those due to the stellar perturbations.

In our previous study~\citep[][here after ZLY15]{Zhang15}, we have developed a fast full general 
relativistic method to obtain the observables of the target star by both considering 
its orbital motion around the MBH and the propagation of photons from the target star to a 
distance observer. We investigated the constraints of the spin parameters by fitting to the 
observables of the target star without considering the stellar perturbations. 
Relying upon the framework of ZLY15, here we further include the gravitational perturbations due 
to a single or a cluster of disturbing object(s) on the orbital motion of the target stars 
orbiting the MBH. By performing a large number of numerical simulations, we investigate the 
Newtonian perturbations on the apparent orbital motion and the redshift of the target star and 
their dependences on the model parameters. The differences between the spin-induced relativistic 
effects and the Newtonian perturbations revealed by our study can provide useful clues of their 
separation methods, which are quite essential for the accurate measurements of the spin parameters 
and also the tests of the Kerr metric.

This paper is organized as follows. Section~\ref{sec:motion_eq} describes the details of the 
numerical methods. The gravitational attractions of the background perturbers are 
included as an additional perturbed Hamiltonian term in the equations of motion of the 
target star. To obtain the projected sky position and the redshift of the star at a given moment, 
we adopt the light tracing technique described in ZLY15 to solve the light trajectories propagating 
from the star to the observer. In Section~\ref{sec:pb_target}, we describe the details of the 
methods used to estimate the perturbations on the apparent position and redshift of the target star 
due to the 
GR spin effects, Newtonian perturbations and their combined effects from the numerical simulations. 
In Section~\ref{sec:nw_single}, we investigate the stellar perturbations caused by a single 
perturber, in the specific case that S2/S0-2 is perturbed by the gravitational force of 
S0-102 (Section~\ref{subsec:S2}), or in the general case that a hypothetical S-star located 
inside the orbits of the current detected ones is perturbed by a single perturber 
(Section~\ref{subsec:inner_Sstar}). In Section~\ref{sec:nw_cluster}, we consider the stellar 
perturbations of a cluster of disturbing objects. By performing a large number of numerical 
simulations, we investigate their resulting perturbations on the observables and comparing them to 
those of the spin-induced signals. The discussions and conclusions are provided in 
Section~\ref{sec:discussion} and Section~\ref{sec:conclusion}, respectively.

\section{Perturbed motion of the target star}
~\label{sec:motion_eq}
The geodesic motion of a star orbiting a Kerr black hole~\citep{Kerr63} 
can be described by a Hamiltonian $H_{\rm K}$. In the Boyer-Lindquist 
coordinates $(r,\theta,\phi, t)$~\citep{Boyer67}, it is given by
\be
\ba
H_{\rm K}=&-\frac{(r^2+a^2)^2-a^2\Delta\rm{sin}^2\theta}{2\Sigma\Delta}p_t^2
-\frac{2ar}{\Sigma\Delta}p_t p_\phi
+\frac{\Delta}{2\Sigma}p_r^2\\
&+\frac{1}{2\Sigma}p_\theta^2
+\frac{\Delta-a^2\sin^2\theta}{2\Sigma\Delta\sin^2\theta}p_\phi^2
~\label{eq:Hk}
\ea
\ee
where  \be\ba
\left\{
\begin{array}{lcl}
\Sigma    & = & r^2+a^2\cos^2\theta, \\
\Delta    & = & r^2-2 r+a^2.\\ 
\end{array}
\right.
\ea\ee
Here $p_t$, $p_r$, $p_\theta$ and $p_\phi$ are the components of the tetrad-momentum 
in the Boyer-Lindquist coordinate. Moreover, $a= Jc/(\bh^2 G)$ is 
the dimensionless spin parameter of the MBH, $J$ is the spin angular momentum of 
the MBH. For simplicity, we set $G = c =\bh = r_\g = G\bh/c^2 
=1$ above, $G$, $c$, $\bh$ and $r_\g$ are the gravitational constant, the speed of light, 
the MBH mass, and the gravitational radius, respectively. Throughout this paper, we 
assume that the MBH in the GC is with a mass of $\bh=4\times10^6\msun$ and a
distance of $R_{\rm GC}=8$ kpc. The corresponding gravitational radius is then given by 
$r_\g\simeq0.04\AU\simeq5\mu$as$\simeq2\times10^{-4}$\,mpc. 

The equations of motion described by $H=H_{\rm K}$ can be further reduced to 
Equation 19-22 in ZLY15, from which we integrate numerically the orbital trajectories of a 
star without any stellar perturbation. As the Hamiltonian relies upon the full Kerr 
metric, all the various GR effects in the orbital motion of the target star around the 
Kerr MBH, including the advancements in the periastron and the orbital 
plane caused by the spin-induced effects, e.g., the frame dragging and the quadrupole effects, are 
thus simultaneously included in the simulations.

If the target star is surrounded by $N_\p$ perturbers, for example, a cluster
of stars or stellar remnants in the field, the gravitational attractions from these sources 
can deviate the orbital motion of the target star from the GR prediction. Considering 
only the leading order perturbations in Newtonian gravity, we ignore the mutual 
gravitational interactions between the perturbers and take the target star as a test 
particle\footnote{We notice that assuming a non-zero mass of the target star could possibly lead 
to nonnegligible back-reaction effects on the motion of the MBH and the perturbers. We defer the exploration 
of such potentially relevant effects for future works.}, then the perturbations on the target star contributed by 
these sources can be approximately expressed by a Hamiltonian $H_{\rm p}$, which is given by~\citep[see 
also][]{Angelil14,Wisdom91},  
\be
\ba
H_{\rm p}&=\sum^{N_\p}_jm_{{\rm p},j}\left(
\frac{r}{r_j^2}\cos\zeta_j-
\frac{1}{d}\right)\\
~\label{eq:Hp}
\ea
\ee
Here $(r_j, \theta_j, \phi_j)$ and $m_{{\rm p}, j}$ is the spatial position and the mass 
of the $j$-th perturber, respectively. 
$\zeta_j=\arccos\left[\sin\theta\sin\theta_j\cos(\phi-\phi_j) 
+\cos\theta\cos\theta_j\right]$ is the angle between the position vector of the target 
star and that of the $j$-th perturber, and $d=\sqrt{r^2+r_j^2-2rr_j\cos\zeta_j}$ is the 
distance between the target star and the $j$-th perturber. Then the motion of 
the perturbed target star can be described by the modified Hamiltonian $H=H_{\rm 
K}+H_{\rm p}$. Note that as the mutual perturbations between the perturbers are 
disregarded, the orbital motion of each perturber rotating around the black hole can be 
fully described by Hamiltonian $H=H_{\rm K}$ and integrated by Equation 19-22 in ZLY15.

The orbital evolutions of the target star and the perturbers can be integrated from
their Hamiltonian equations of motion once the initial orbital elements of the star and of the 
perturbers, and also the mass, the spin parameters and the distance of the MBH are provided~(see 
details in ZLY15). Here the six orbital elements of the target star (the perturber) are the 
semimajor axis $a_\star$ ($a_{\rm p}$), eccentricity $e_\star$ ($e_{\rm p}$), inclination $I_\star$ 
($I_\p$), longitude of position angle of the ascending 
node $\Omega_\star$ ($\Omega_\p$), arguments of angle to periapsis $\omega_\star$ 
($\omega_\p$), and the true anomaly $f_\star$ 
($f_\p$) [or the time of pericenter passage $t_{0\star}$ ($t_{0\rm p}$)], which are defined 
respective to the sky plane. The spin direction of the MBH is defined by two angles: $i$ and 
$\epsilon$. Here $i$ is the line of sight inclination of the spin, $\epsilon$ is the angle between 
the projection of the direction of the spin onto the sky plane and a reference 
direction~\footnote{For the definitions of the orbital elements of the star and the spin angles of 
the MBH please see the Figure 1 in ZLY15, identifying $a_\star=a_{\rm orb}'$, $e_\star=e_{\rm 
orb}'$, $I_\star=I'$, $\Omega_\star=\Omega'$, $\omega_\star=\Upsilon'$ and $f_\star=\nu'$ for 
$a_{\orb}'$, $e_{\orb}'$, $I'$, $\Omega'$, $\Upsilon'$ and $\nu'$ defined in ZLY15.}.

The observed sky position and the redshift of the target star can be obtained if the 
parameters describing the light trajectory emitted from the star to the distant 
observer are determined. As the spin-induced position difference of the star in the sky 
plane are found approximately the same order of magnitude compared to those caused by light-bending 
effects, it is crucial to include the light-tracing technique in the simulations to 
model accurately the apparent position of the star (See Section 5.1 in ZLY15). As the total  
enclosed mass of the perturbers distributed around the MBH (typically $\la 10$mpc in this paper, 
see also Section~\ref{sec:nw_cluster}) are much smaller than the mass of the central MBH, we assume 
that the light trajectories can be approximately integrated by the equations of motion in the Kerr 
metrics. In this work, we adopt the backward light-tracing technique described in ZLY15 to calculate 
the observed right ascension (R.A.), the declination (Dec) of the target star in the sky plane, and 
also the redshift ($Z$) of the target star as a function of the 
observational time $t_{\rm obs}$ for an observer located at distance $R_{\rm GC}$.
Here $t_\obs=t_\star+t_{\rm prop}$, $t_\star$ is the local time of the star in the Boyer-Lindquist 
coordinates and $t_{\rm prop}$ is the time used for a photon propagating from the star to the 
observer. All of the various GR effects affecting the propagations of the photons from the star to 
the observer, including both the displacement of the image position in the sky plane due to the 
gravitational light-bending and the gravitational redshift of the 
target star~(e.g.,\citealt{Iorio11b,Angelil10a}; ZLY15), are then simultaneously included in the 
mock observables of the target star.

The full GR effects can be divided into two parts: the spin-zero  and the spin-induced effects. The 
spin-zero effects are those when the black hole is not spinning, e.g., the Schwarzschild 
precession, the time dilation, and etc. The spin-induced effects include the frame-dragging ($\propto a$), quadrupole 
momentum ($\propto a^2$), and other high order spin-related GR effects as well. Both of the spin-zero and the 
spin-induced effects have been automatically included in the orbital motion and the mock observables of 
the target star as we adopt a full Kerr metric. However, in the following sections, we deal only with the 
spin-induced effects, and single them out by removing the spin-zero effects from the full GR effects. 
For a given set of initial conditions, this can be done by examining the differences between the results of the simulation 
with $a=0.99$ and that with $a=0$ (See details in Section~\ref{sec:pb_target}). We will not discuss about the effects of the 
spin-zero terms as they does not have direct connections to the purpose of this study. We defer the discussions 
of the spin-zero effects and their differences with the Newtonian perturbations to future works. 

The Keplerian orbital elements of a star can be obtained by its instantaneous position and velocity. 
In this work, we calculate the orbital elements of the target star at any given moment by its three-position and three-velocity 
measured in the local non-rotating rest frame (LNRF) of the target star. Here the three-velocity can be derived 
according to Eq. 9-11 of ZLY15. We notice that if alternatively the elements are estimated by instantaneous position and the
momentum in the LNRF frame~\citep[see also][]{PS09}, the results will be slightly different. We also notice that the orbital elements
calculated in this work can be different with those obtained in the post-Newtonian (PN) simulations, as the adopted spacetime metrics are different. However, we find that the results of these two methods are generally consistent (See more details in Section~\ref{subsubsec:com_pn}).

\section{Evaluating the perturbations felt by the target star}
~\label{sec:pb_target}
The orbital motion and the observables of a target star rotating around a spinning black hole can 
be affected by both the spin-induced effects and the gravitational attractions from other 
stars/stellar remnants. In this work, we mainly deal with three types of perturbations 
experienced by the target star: (1) The spin-induced perturbation; (2) The Newtonian perturbation; 
(3) The combination of (1) and (2).  If $Y$ is any quantity relative to the motion of the 
target star, then the three types of perturbations listed above can be denoted as $\delta_\s Y$, 
$\delta_\p Y$ and $\delta_\c Y$, respectively. Here, $Y$ can be any one of the orbital elements of the target star, e.g., 
$a_\star$, $e_\star$, $I_\star$, $\Omega_\star$, $\omega_\star$, $f_\star$, $\cdots$, or the 
observables, e.g., the coordinates of the apparent position (${\rm R.A.}$ and ${\rm Dec}$) and the 
redshift ($Z$) of the star. Note that $Y$ can also be the position vector of the target star appeared 
in the sky plane, i.e., $\mathbf{R}=($ R.A., Dec$)$; then, its perturbations are denote as $\delta 
\mathbf R=(\delta{\rm R.A.}, \delta{\rm Dec})$.

The three types of the perturbations can be obtained by comparing between two simulations that 
have the same set of the initial orbital parameters of the target star and the perturber(s), 
however, the effects of spin and/or Newtonian gravities are turned on in one simulation and off 
in the other: (1) $\delta_\s Y$ is the difference of $Y$ in the simulation with $a=0$ and that with
$a=0.99$, both of which ignore the stellar perturbations;  (2) $\delta_\p Y$ is the difference of $Y$ 
in the simulation ignoring and that including the Newtonian gravities of the perturbers, both 
of which the spin is set to $a=0$. (3) $\delta_\c Y$ is the difference of $Y$ between the simulation including both the 
Newtonian and spin-induced perturbations, and that ignoring both of them. We found that 
$\delta_\c Y\simeq\delta_\p Y+\delta_\s Y$.
 
As showed in the following sections, the signals of all these perturbations usually 
vary significantly as a function of time. The overall contributions of these perturbations can be 
estimated by their root mean squared values. Within a period of time $T_{\rm tot}$, the root mean 
squared (RMS) perturbation in $Y$ is defined by
\be
\ba
\overline{\delta Y}=\sqrt{\frac{ 1}{T_{\rm tot}}\int^{T_{\rm tot}}\delta 
Y(t_\obs)^2dt_\obs}.
\label{eq:yrms}
\ea
\ee

Note that for the apparent position vector of the target star, i.e., $\mathbf{R}=($R.A., Dec$)$, the RMS magnitude of its perturbation, i.e., 
$|\delta\mathbf{R}|=\sqrt{\delta{\rm Dec}^2+ \delta{\rm R.A.}^2}$, is defined by 
\be
\overline{|\delta \mathbf {R}|}=\sqrt{\frac{ 1}{T_{\rm tot}}\int^{T_{\rm tot}}|\delta 
\mathbf{R}(t_\obs)|^2dt_\obs}.
\label{eq:rrms}
\ee
Similarly, if ``$\delta$'' before $Y$ or $\mathbf{R}$ in Equation~\ref{eq:yrms} or 
\ref{eq:rrms} is replaced by ``$\delta_\s$'', ``$\delta_\p$'', or ``$\delta_\c$'', it 
means the RMS spin-induced perturbations, stellar perturbations or the 
combinations of them, respectively.

\section{The Newtonian perturbations of a single perturber}
~\label{sec:nw_single}
\begin{figure}
\center
\includegraphics[scale=0.70]{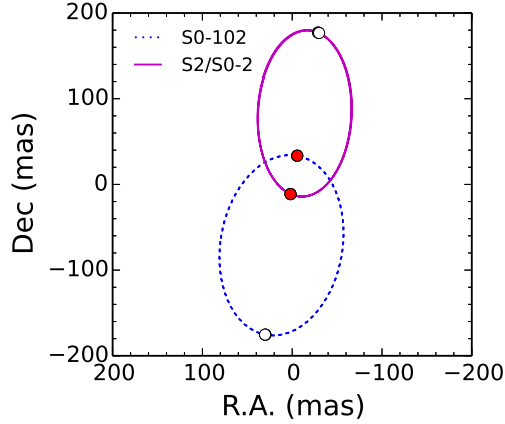}
\caption{Apparent trajectories of S2/S0-2 (solid magenta line) and S0-102 (the dotted blue line)
in the sky plane over three orbits. Red solid and white open circles mark the locations of the 
periapsis and apoapsis, respectively.
}
\label{fig:img}
\end{figure}
\begin{figure*}
\center
\includegraphics[scale=0.60]{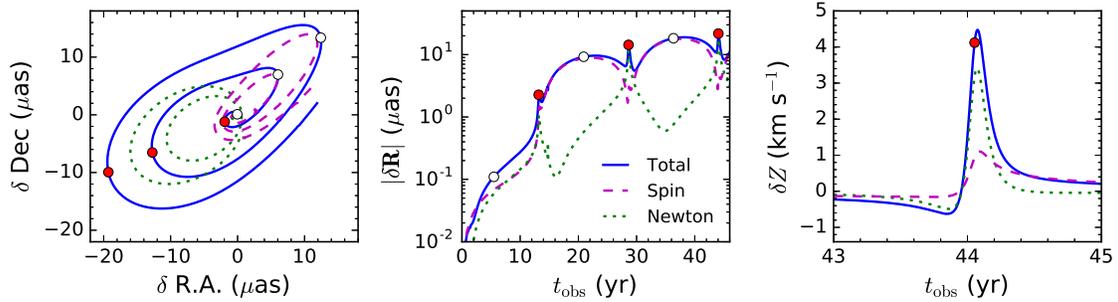}
\caption{Perturbations on the apparent position [$\delta \mathbf{R}=(\delta{\rm R.A.},
\delta{\rm Dec})$, left panel], its distance in the sky plane ($|\delta \mathbf{R}|$, 
middle panel), and the redshift ($\delta Z$, right panel) of S2/S0-2 as a function of the 
observational time ($t_\obs$) in three orbits. Note that in the right panel, only the evolutions of 
$\delta Z$ near the third pericenter passage of S2/S0-2 ($t_\obs\simeq 44.1$ yr) is plotted. In 
each panel, the perturbations on these signals of star S2/S0-2 are due to the spin-induced effects 
when $a=0.99$, $i=45^\circ$ and $\epsilon=200^\circ$ (magenta dashed lines), the Newtonian 
attractions of S0-102 with $m_\p=0.5\msun$ (green dotted lines), and the combination effects of the 
above two (blue solid lines), all of which are obtained by the method described in 
Section~\ref{sec:pb_target}. Red solid and white open circles mark the periapsis 
and apoapsis passage points of S2/S0-2, respectively.
}
\label{fig:sg_S2}
\end{figure*}
\begin{figure}
\center
\includegraphics[scale=0.8]{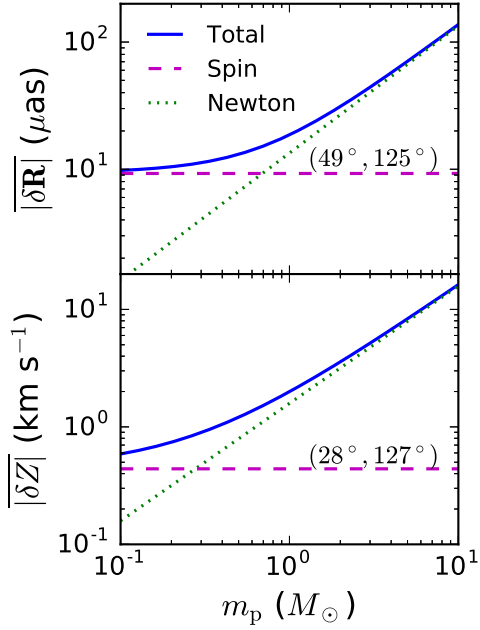}
\caption{RMS values of the perturbations on the position signal ($\overline{|\delta 
\mathbf{R}|}$, top panel) or the redshift signal ($\overline {|\delta Z|}$, bottom panel) of 
S2/S0-2 over three orbits versus the mass of S0-102 ($m_\p$). In each panel, the perturbations are 
due to the spin-induced effects (magenta dashed line), the Newtonian attractions of S0-102 
(green dotted line), and the combination effects of the above two (blue solid line). In the 
top panel, the spin orientation is $i=49^\circ$ and $\epsilon=125^\circ$, such that the 
spin-induced position displacement of S2/S0-2 are most significant; In the bottom panel, the spin 
orientation is $i=28^\circ$ and $\epsilon=127^\circ$, such that the spin-induced redshift 
difference of S2/S0-2 are most significant.
}
\label{fig:mass_S0-102}
\end{figure}
\begin{figure*}
\center
\includegraphics[scale=0.60]{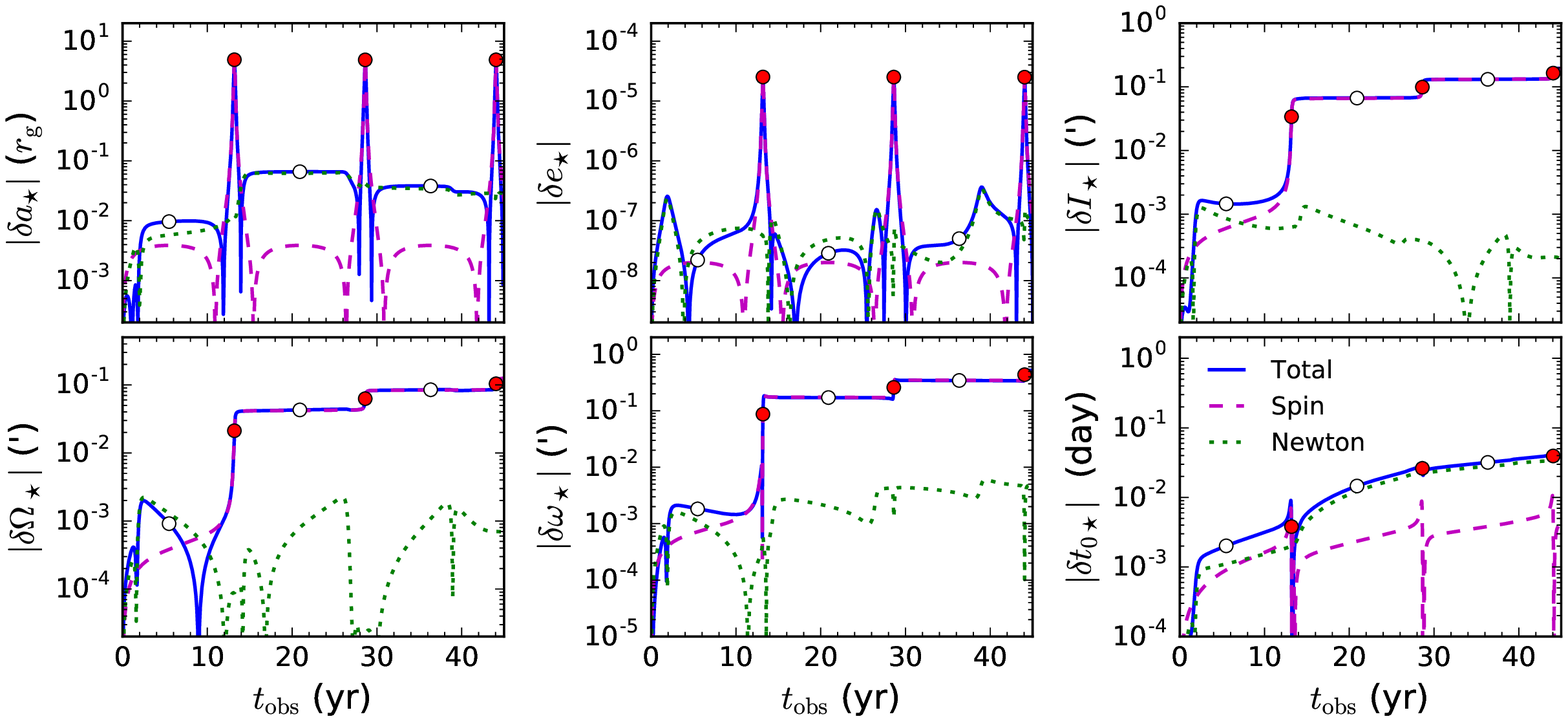}
\caption{Perturbations on the six orbital elements of S2/S0-2, which are the orbital 
semimajor axis ($a_{\star}$, top left panel), eccentricity ($e_{\star}$, top middle panel), 
inclination ($I_\star$, top right panel), ascending node ($\Omega_\star$, bottom left panel), 
longitude of angle to periapsis ($\omega_\star$, bottom middle panel), and the time of 
pericenter passage 
($t_{0\star}$, bottom right panel). In each panel, the perturbations are 
due to the spin-induced effects when $a=0.99$, $i=45^\circ$ and $\epsilon=200^\circ$(magenta dashed 
line), the Newtonian attractions of S0-102 with $m_\p=0.5\msun$ (green dotted line), and the 
combination effects of the above two (blue solid line). Red solid and white open
circles mark the periapsis and apoapsis passage points of S2/S0-2, respectively.
}
\label{fig:ele_S2}
\end{figure*}

In this section, we study the simple case that the target star orbiting the MBH is perturbed 
by a single perturber. The target star and the perturber are considered to be the currently 
detected ones (See Section~\ref{subsec:S2}), or those undetected, but very likely to be existed in 
the vicinity of the MBH in the GC (See Section~\ref{subsec:inner_Sstar}). The importance of 
such analysis is mainly threefold: (1) As we will see in the latter sections, such studies help 
to understand the behaviors of the Newtonian perturbation and its difference with the spin-induced 
effects; (2) They help to reveal the Newtonian perturbations contributed by the individual 
perturber when the target star is embedded in a stellar cluster. (3) As currently there are 
relatively large uncertainties for the mass profiles in the vicinity of the MBH (See details in 
Section~\ref{sec:nw_cluster}), it remain possible that a target star within a few hundred AU from 
the MBH is attracted by just a few perturbers located inside or nearby (See also 
Figure~\ref{fig:profile}). In these cases, such a target star-perturber-MBH three body problem may 
be important if one of the perturbers among them play a dominant role in the signals of the 
Newtonian perturbations. 

\subsection{Perturbations on S2/S0-2 from S0-102}
~\label{subsec:S2}

\begin{table}
\caption{Initial Orbital Elements of the Target stars and Perturbers.}
\centering
\begin{tabular}{lccccccccc}
\hline
\multirow{2}{0.5cm}{Target star} && $a_\star$ &$r_{{\rm per},\star}$ $^{a}$ & $e_\star$ & $I_\star$  & 
$\Omega_\star$ & $\omega_\star$ & ${t_{0\star}}^{b}$ & $f_\star$ 
 \\
 && ($\AU$) & ($r_\g$) & & ($^\circ$) & ($^\circ$) & ($^\circ$) & (yr) & ($^\circ$) \\
\hline 
S2/S0-2$^{c}$ && 984 &$2993$ & 0.88 & 135 & 225 & 63 & 2.32 & --\\
T1$^{e}$ && $50-800$ & $152-2434$ & 0.88 & 45 & 0 & 0 & -- & 180\\
T2$^{e}$ && $50-800$  &$887-14198$ & 0.3 & 45 & 0 & 0 & -- & 180\\
\hline
\multirow{2}{0.5cm}{Perturber}& & $a_\p$ & $r_{{\rm per},\p}$ $^{a}$ & $e_\p$ & $I_\p$  & 
$\Omega_\p$ & $\omega_\p$ & ${t_{0\p}}^{b}$ & $m_\p$\\
& & ($\AU$) & ($r_\g$) & & ($^\circ$) & ($^\circ$) & ($^\circ$) & (yr) & $(\msun)$ \\
\hline
S0-102$^{d}$ & & 848  &$6880$& 0.68 & 151 & 175 & 185 & 9.5 & $0.5$\\
P1$^{f}$ && 100  & $304$ & 0.88 & Ran & Ran & Ran & Ran & $10$  \\
P2$^{f}$ && 600  & $1825$ & 0.88 & Ran & Ran & Ran &  Ran & $10$ \\
\end{tabular}
\label{tab:t1}
\tablecomments{\\
\, $^a$ Distance at pericenter.\\
\, $^b$ Time of the pericenter passage respective to the year of 2000.\\
\, $^c$ Taken from~\citet{Gillessenetal09}.\\
\, $^d$ Taken from~\citet{Meyer12}. \\
\, $^e$ The orbital semimajor axis ($a_\star$) adopts one of the ten 
values that are logarithmically spaced between $50\AU$ and $800\AU$, which are $50\AU$, $68\AU$, 
$93\AU$, $126\AU$, $171\AU$, $233\AU$, $317\AU$, $432\AU$, $588\AU$, and $800\AU$.
The distance at pericenter ($r_{{\rm per},\star}$ in unit of $r_\g\simeq0.04\AU$) of T1 
and T2 can then be obtained correspondingly.\\
\, $^f$ In the MC simulation, $I_\p$, $\Omega_\p$, $\omega_\p$ of the perturber are
with random values between $0^\circ$ and $360^\circ$, $t_{0\p}$ of the perturber is with random 
values between $0$ and $P(a_\p)$, here $P(a_\p)$ is the orbital period of the perturber.
}
\end{table}

Among all of the currently detected S-stars, the star S2/S0-2 is of particular interests 
for testing the spin-induced effects. As it is in close proximity to the MBH and 
have relatively high eccentricity, the continuous monitoring of the orbital motion 
of S2 by the future facilities, i.e., TMT or E-ELT, can be used to provide constraints on the 
spin parameters of the MBH (See ZLY15). However, it requires a clean separation of the stellar 
perturbations caused by other surrounding S-stars or the still undetected stars/stellar remnants. 
One of the closest S-stars is the recently discovered S0-102, which has the shortest orbital 
period known so far~\citep{Meyer12}. It has been found that S0-102 is not particularly suited to 
probe the spin parameters of the MBH (See ZLY15), but it may introduce gravitational 
perturbations on the orbits of S2/S0-2. In this section, we study the perturbations on 
both the orbit elements and the direct observables of S2/S0-2 due to the gravitational pull from 
S0-102.

We adopt the initial conditions of the target star S2/S0-2 and the perturber S0-102 
according to the recent observations~\citep[][see 
Table~\ref{tab:t1}]{Gillessenetal09,Meyer12}. We simulate the orbital motion of S2/S0-2 over 
three orbital periods, beginning in the year of 2020 (corresponding to $t_{\rm obs}=0$), to mimic 
the observational signals probed by the future telescopes. As currently the mass of S0-102 is 
poorly known, we assume $m_\p=0.5\msun$ for it. As we will see later in this section, such assumed 
mass of S0-102 result in Newtonian perturbations almost comparable with those caused by the 
spin-induced effects on both the position and redshift signals of the S2/S0-2. We will discuss the 
dependence of the results on the mass of S0-102 later in this section.

For the magnitude and direction of the MBH spin we assume three different cases: (1)
$a=0.99$, $i=45^\circ$ and $\epsilon=200^\circ$, such that both the spin-induced effects on the 
position and redshift signals of S2/S0-2 are modest; (2) $a=0.99$, $i=49^\circ$ and 
$\epsilon=125^\circ$, such that the spin-induced position displacement of S2/S0-2 are most 
significant. (3) $a=0.99$, $i=28^\circ$ and $\epsilon=127^\circ$, such that the spin-induced 
redshift differences of S2/S0-2 are most significant~\citep[See also][]{Yu16}. 

\subsubsection{Perturbations on the observables of S2/S0-2}
~\label{subsubsec:pb_ob_S2}

The simulated apparent trajectories of these two stars in the sky plane are showed in 
Figure~\ref{fig:img}. Figure~\ref{fig:sg_S2} show the simulation results of the perturbations 
on the observables of S2/S0-2 due to the spin-induced effects when $a=0.99$, $i=45^\circ$ and 
$\epsilon=200^\circ$ (dashed magenta lines), the Newtonian attractions of S0-102 (green dotted 
lines), and the combination effects of them (solid blue lines). Table~\ref{tab:t2} shows the 
maximum and the RMS values of these perturbations over the three orbits. The details of the results 
are discussed as follows.

The evolutions of the spin induced perturbations on the apparent position ($\delta_\s\mathbf{R}$), 
its distance in the sky plane ($|\delta_\s \mathbf{R}|$), and the redshift ($\delta_\s Z$) of 
S2/S0-2 are showed in the dashed magenta lines in left, middle and right panel 
of Figure~\ref{fig:sg_S2}, respectively. We can see that the spin-induced position displacement is 
spiral-like, mounts up and peaks near the apocenter in each orbit, while the spin-induced 
redshift difference is most variable and peaks near the pericenter in each orbit.
Note that these spin-induced effects depend on the assumed spin orientations. The values of the 
maximum and RMS spin-induced perturbations over three orbits in three cases of  
spin orientations are showed in Table~\ref{tab:t2}. 

The Newtonian perturbations on both the position and redshift signals of S2/S0-2 show quite 
different evolutions compared with those of the spin-induced effects (See 
the dotted green lines in Figure~\ref{fig:sg_S2}). The Newtonian perturbation on both position 
and redshift signals peak around the pericenter in each orbit, with the 
maximum values given by $|\delta_\p \mathbf{R}|_\max\sim 17.9\mu$as and $|\delta_\p 
Z|_\max\simeq3.4\kms$ around the third pericenter passage. The corresponding RMS values are 
given by $\overline{|\delta_\p\mathbf{R}|}=6.7\mu$as and $\overline{\delta_\p Z}=0.8\kms$ for the 
apparent position and the redshift signals, respectively. 

The combined perturbations of the Newtonian and the spin-induced effects are 
complex (See the blue solid lines in each panel of Figure~\ref{fig:sg_S2}), as in the simulations 
here they are comparable to each other, i.e., $\overline{|\delta_\p 
\mathbf{R}|}\simeq\overline{|\delta_\s\mathbf{R}|}$ or $\overline{|\delta_\p 
Z|}\simeq\overline{\delta_\s Z}$. Note that the combined perturbations on the position signal of 
S2/S0-2 show peaks both near the pericenter and apocenter, which are contributed by the 
Newtonian and the spin-induced effects, respectively.

\begin{landscape}
\begin{table*}
\caption{Perturbations on position and redshift of S2/S0-2 over three orbital periods}
\centering
\begin{tabular}{cccccccccccccccccc}
\hline\\[-8pt]
$(i,\epsilon)$& 
$|\delta_\s \mathbf{R}|_\max$&
$|\delta_\p \mathbf{R}|_\max$&
$|\delta_\c \mathbf{R}|_\max$& 
$\overline{|\delta_\s \mathbf{R}|}$  &
$\overline{|\delta_\p \mathbf{R}|}$  &
$\overline{|\delta_\c \mathbf{R}|}$  &
$|\delta_\s Z|_\max$  &
$|\delta_\p Z|_\max$  &
$|\delta_\c Z|_\max$  &
$\overline{|\delta_\s Z|}$  &
$\overline{|\delta_\p Z|}$  &
$\overline{|\delta_\c Z|}$  & 
$m^{\rm zc}_\p$ &
$m^{\rm rc}_\p$  \\
$[1]$ & [2]&[3]&[4]&[5]&[6]&[7]&[8]&[9]&[10]&[11]&[12]&[13]&[14]&[15]\\[2pt]
\hline\\[-7pt]
($45^\circ, 200^\circ$)    & 
17.8   &  17.9 & 22.3  &
6.1 & 6.7& 10.5&
1.1 & 3.4& 4.4 &
0.32 & 0.79 & 1.1 & 
0.45 & 0.20
\\
($49^\circ, 125^\circ$)    & 
26.5   &  17.9 & 27.2 & 
9.3 & 6.7& 13.1&
1.5 & 3.4& 4.9& 
0.41 & 0.79 & 1.2 &
0.69 & 0.26
\\
($28^\circ, 127^\circ$)    & 24.8   &  17.9 & 25.7 &  
8.5 & 6.7& 12.5&
1.6 & 3.4& 5.0& 
0.44 & 0.79 & 1.2 & 0.64 & 0.28
\\
\end{tabular}
\label{tab:t2}
\tablecomments{Summaries of the perturbations on position and redshift signals of S2/S0-2 
over three orbital periods. Col.[1]: $i$ and $\epsilon$ are the two angles defining the spin 
orientation: $i$ is the line of sight inclination of the spin, $\epsilon$ is the angle between the 
projection of the direction of the spin onto the sky plane and a reference direction~(See Figure 1 
in ZLY15). Col.[2-7]: The maximum (Col.[2-4]) and RMS values (Col.[5-7]) 
of the perturbations on the position signal of S2/S0-2 over three orbits, in units of 
$\mu$as. Col.[8-13]: The maximum (Col.[8-10]) and RMS values (Col.[11-13]) of the perturbations 
on the redshift signal of S2/S0-2 over three orbits, in units of $\kms$. In Col.[2-13], 
the quantities with ``$\delta_\s$'', ``$\delta_\p$'', and ``$\delta_\c$'' mean the perturbations 
due to spin-induced effects, the Newtonian attractions of S0-102 with $m_\p=0.5\msun$, and 
the combination effects of them, respectively. Col.[14-15]: $m_\p^{\rm rc}$ is the mass of 
S0-102 when $\overline{|\delta_\p \mathbf{R}|}=\overline{|\delta_\s \mathbf{R}|}$; $m_\p^{\rm zc}$ 
is the mass of S0-102 when $\overline{|\delta_\p Z|}=\overline{|\delta_\s Z|}$.
}
\end{table*}
\end{landscape}

We find that the Newtonian perturbations are proportional to the mass of S0-102 ($m_\p$), as 
showed in the dotted green lines in Figure~\ref{fig:mass_S0-102}. The RMS Newtonian perturbation 
are given by $\overline{|\delta_\p \mathbf{R}|}\simeq 13.4\,(m_\p/1\msun) \mu$as 
and $\overline{\delta_\p Z}\simeq 1.6\,(m_\p/1\msun)\kms$ for the position and redshift signals of 
S2/S0-2, respectively. Thus, there is a critical mass of S0-102, i.e., $m_{\rm p}^{\rc}$ (or 
$m_{\rm p}^{\zc}$), when the RMS value of the Newtonian perturbations on position (or redshift) 
signal is equal to those resulting from the spin-induced effects, i.e., $\overline{|\delta_\p 
\mathbf{R}|}=\overline{|\delta_\s \mathbf{R}|}$ (or $\overline{\delta_\p Z}=\overline{\delta_\s 
Z}$). The spin effects in position (or redshift) of S2/S0-2 will be obscured by the Newtonian 
perturbations if $m_\p>m_{\rm p}^{\rc}$ (or $m_\p>m_{\rm p}^{zc}$). The critical 
masses in the cases of different spin orientations are showed in Table~\ref{tab:t2}. When the 
spin-induced effects become most significant, we get the upper limit of the critical mass, i.e., 
$m_{\rm p}^{\rc}=0.7\msun$ and $m_{\rm p}^{\zc}=0.28\msun$, for the position and redshift signals 
of S2/S0-2, respectively. As the S-stars are exclusively B-type main-sequence stars with masses 
$\ga3\msun$~\citep{Ghezetal08,Gillessenetal09}, S0-102 is likely more massive than these critical 
values. Thus, it is quite plausible that the spin-induced signals of S2/S0-2 are obscured by the 
perturbations from S0-102 alone.

Observationally, the spin-induced effects can be measured by subtracting the stellar 
perturbations from the total measured shifts. In the case of S0-102, they may be well removed in 
the ideal case when it is the only S-star closer to the MBH than S2/S0-2, and if the orbital 
parameters and the mass of S0-102 are observationally determined. The difference between the
spin-induced effects and the stellar perturbations from S0-102 may provide helpful hints 
on discerning them from the observational data.

\subsubsection{Perturbations on the orbital elements of S2/S0-2}
~\label{subsubsec:pb_orb_S2}
The standard osculating orbital elements of stars are instantaneously calculated from the values of 
their position and velocity vectors from usual Keplerian formulas. Here we calculate the orbital 
elements of S2/S0-2 by the three-position and the three-velocity in its LNRF frame (See Eq. 9-11 in ZLY15). 
Both the Newtonian perturbations from S0-102 and the relativistic spin-induced effects induce variations of 
the orbital elements of S2/S0-2. The perturbations on the observables, i.e., 
$\delta \mathbf{R}$ and $\delta Z$, are approximately related to the variations of the orbital 
elements by 
\be
\delta 
\mathbf{R}(t_\obs)\simeq\sum_\kappa \frac{\partial \mathbf{R}}{\partial \kappa} \delta 
\kappa(t_\obs),
\label{eq:rt}
\ee
and
\be 
\delta Z(t_\obs)\simeq\sum_\kappa \frac{\partial Z}{\partial \kappa} \delta \kappa(t_\obs),
\label{eq:zt}
\ee 
respectively, where $\kappa=a_\star$, $e_\star$, $I_\star$, $\Omega_\star$, 
$\omega_\star$ and $t_{0\star}$ (or $f_{\star}$). We calculate the 
perturbations on the orbital elements of S2/S0-2 according to the method described in 
Section~\ref{sec:pb_target}. The results in the case that $a=0.99$, $i=45^\circ$ and 
$\epsilon=200^\circ$ are depicted in Figure~\ref{fig:ele_S2} and discussed as follows (The results 
for other two spin orientations are similar).

The spin-induced effects cause the variations of all the orbital elements (See the 
magenta dashed lines in Figure~\ref{fig:ele_S2}). The perturbations on the semimajor axis 
$|\delta_\s a_\star|$ and eccentricity $|\delta_\s e_\star|$ due to the spin-induced 
effects oscillate periodically, with local maximums at both periapsis and apoapisis. The 
orbit-averaged values of $\delta_\s a_{\star}$ and $\delta_\s e_\star$ remain constant as a 
function of time. Due to the frame-dragging and also other high order GR effects, $\delta_\s 
I_\star$, $\delta_\s \Omega_\star$, and $\delta_\s \omega_\star$ increase with time. At the third 
pericenter, the variations of these orbital elements amount to $\delta_\s I_\star=0.16'$, 
$\delta_\s \Omega_\star=-0.10'$, and $\delta_\s\omega_\star=-0.44'$. The time of pericenter passage 
$\delta_\s t_{0\star}$ (or similarly, $\delta_\s f_\star$) increases also slightly at each orbit, 
as the orbital period of the star orbiting a MBH with, and without spinning, is different. 
One can show that the spin-induced position difference $|\delta_\s \mathbf{R}|$ and 
$\delta_\s Z$ are contributed by both the orbital precession ($\delta_\s I_\star$, $\delta_\s 
\Omega_\star$ and $\delta_\s\omega_\star$) and also the $\delta_\s t_{0\star}$ (or $\delta_\s 
f_\star$)~(See also~\citealt{Yu16}; ZLY15).

The gravitational attractions of S0-102 can also cause changes of all the orbital elements of 
S2/S0-2 (See the green dotted lines in Figure~\ref{fig:ele_S2}). We find that the stellar 
perturbations on the orbital elements are complex and quite different from the spin-induced 
relativistic ones. Particularly, we found that the orbit-averaged Newtonian perturbations on the 
orbital semimajor axis do not remain constant but vary after each orbit. At the third pericenter, 
the variations of the elements of S2/S0-2 are given by 
$\delta_\p a_\star\simeq0.03r_\g$, $\delta_\p e_\star\simeq\uwave{10^{-7}}$, 
$\delta_\p I_\star\simeq0.014''$, $\delta_\p\Omega_\star\simeq0.041''$, 
$\delta_\p\omega_\star\simeq-0.0035''$ and $\delta_\p 
t_{0\star}\simeq0.034$day (or $\delta_\p f_\star=-4.3'$). 
As $\delta_\p I_\star \ll\delta_\s I_\star$, $\delta_\p\Omega_\star\ll\delta_\s 
\Omega_\star$, $\delta_\p \omega_\star\ll\delta_\s\omega_\star$, the Newtonian orbital precessions 
are negligible compared to the relativistic spin-induced ones. 
As showed by the analytical calculations below, it turns out that the Newtonian perturbations on 
the observables, $|\delta_\p \mathbf{R}|$ and $\delta_\p Z$, are mainly explained by perturbed 
orbital periods (or $\delta_\p f_\star$), rather than the Newtonian orbital precessions.

The changes of the orbital semimajor axis, i.e., $\delta_\p a_\star$, due to the Newtonian 
attractions of S0-102, cause variations of the orbital period and the time of 
arrival in each point in the orbit of the target star. Suppose that the evolution of the true 
anomaly of the S2/S0-2 with and without perturbations of S0-102 are given by $f_\star'(t_\star')$ 
and $f_\star(t_\star)$, respectively, then when the star reach to the same true anomaly, i.e., 
$f'(t_\star')=f(t_\star)$, the difference of the time of arrival in these two cases is given by
$\delta_\p t_\star=t'_\star-t_\star$. If we assume that the perturbation is small, i.e., 
$df'(t_\star)/dt_\star \simeq df(t_\star)/dt_\star$, then at a given moment $t_\star$, the 
difference of the true anomaly is given by 
\be\ba
\delta_\p f_\star(t_\star)&=f_\star'(t_\star)-f_\star(t_\star)\\
&\simeq-\frac{df_\star}{dt_\star}\delta_\p 
t_\star\simeq-\frac{\sqrt{\bh G a_\star (1-e_\star^2)}}{r^2}\delta_\p t_\star.
~\label{eq:dp_f}
\ea
\ee

At the pericenter, we simply have $\delta_\p t_\star=\delta_\p t_{0\star}$. Then the variation of 
the true anomaly is given by
\be\ba
 \delta_\p f_\star \simeq- &\frac{v_{\rm per}}{a_\star(1-e_\star)} \delta_\p 
t_{0\star} .
~\label{eq:df_f_per}
\ea\ee
Here $v_{\rm per}=\sqrt{\frac{\bh G}{a_\star}} 
\sqrt{\frac{1+e_\star}{1-e_\star}}$ is the velocity of the star at the pericenter. Note that if 
substituting the simulation result $\delta_\p t_{0\star}=0.034$day to 
Equation~\ref{eq:df_f_per}, we have $\delta_\p f_\star=-4.3'$, 
which is also consistent with the result of the 
simulation.

At the third pericenter, the resulting position displacement and difference of the velocity in the 
line of sight (which can be approximately regarded as the redshift) due to the perturbed period (or 
the variations of $f_\star$) are given by
\be
\ba
|\delta_\p 
\mathbf{R}|_{f_\star=0}&\simeq a_\star(1-e_\star)
(1-\cos^2\omega_\star\sin^2I_\star)^{1/2}|\delta_\p f_\star|\\
&\simeq17.8\mu{\rm as}
~\label{eq:r}
\ea
\ee
and
\be\ba
|\delta_\p Z|_{f_\star=0}&\simeq\sqrt{\frac{G\bh}{a_\star(1-e_\star^2)}}\sin\omega_\star\sin 
I_\star|\delta_\p f_{\star}|\\
&\simeq3.2\kms,
~\label{eq:z}
\ea
\ee
respectively. This analysis is well consistent with the numerical simulation results, i.e., 
$|\delta_\p \mathbf{R}|\simeq 17.9\mu$as and $|\delta_\p Z|\simeq 3.4\kms$, at the third pericenter 
of S2/S0-2 (See Table~\ref{tab:t2}). 

Similarly, we can obtain the analytical expressions of 
$|\delta_\p\mathbf{R}|$ and $|\delta_\p Z|$ for arbitrary $f_\star$. Note that according to 
Equation~\ref{eq:dp_f}, $\delta_\p f_\star$ peaks around the pericenter in each orbit, thus the 
resulting perturbations on the observables of S2/S0-2 also peak around the pericenter (See 
Section~\ref{subsubsec:pb_ob_S2} and Figure~\ref{fig:sg_S2}). 

Meanwhile, we find that the Newtonian orbital precessions induce negligible changes on the 
observables of S2/S0-2. According to Equation~\ref{eq:rt} and~\ref{eq:zt}, near the third 
pericenter passage, the changes of the observables due to the Newtonian orbital precessions are 
given by (Similar to Equation 31 and 36 in ZLY15)
\be\ba
|\delta_\p \mathbf{R}|_{\rm prec} &\simeq a_\star (1-e_\star) 
\left[\delta_\p^2 \Omega_\star(1-\sin^2\omega_\star \sin^2 I_\star)\right.\\
&+\delta_\p^2 \omega _\star(1-\cos^2\omega_\star\sin^2 I_\star)\\
&+\sin^2\omega_\star\sin^2 I_\star\delta_\p^2 I_\star+ 2\cos I_\star\delta_\p 
\Omega_\star\delta_\p\omega_\star\\
&-2\sin\omega_\star\cos\omega_\star\sin I_\star \cos I_\star\delta_\p\omega_\star\delta_\p I_\star\\
&\left.-2\sin\omega_\star\cos\omega_\star\sin I_\star \delta_\p\Omega_\star\delta_\p I_\star
\right]^{1/2}\\
&\simeq 0.03\mu{\rm as}
~\label{eq:ar}
\ea
\ee
and 
\be\ba
|\delta_\p Z|_{\rm prec} &\simeq \sqrt{\frac{G\bh}{a_\star(1-e_\star^2)}} (1+e_\star) 
\left(\cos\omega_\star \cos I_\star\delta_\p I_\star\right.\\
&\left.-\sin\omega_\star \sin I_\star\delta_\p \omega_\star\right)\\
&\simeq10^{-3}\kms,
~\label{eq:az}
\ea
\ee
for the position and redshift signals, respectively. These values are orders of magnitude lower 
than those in Equation~\ref{eq:r} and~\ref{eq:z}. 

Similarly, for most of the simulations performed in the latter sections, i.e., in 
which a generic target star is perturbed by a single (Section~\ref{subsec:inner_Sstar}) or multiple 
disturbing objects (Section~\ref{sec:nw_cluster}), we find that the Newtonian perturbations on the 
observables are mainly caused by the perturbed orbital periods of the target star, rather than the 
Newtonian orbital precessions. Only in some rare cases, that the perturber is with some particular 
orbital configurations, the effects due to the perturbed orbital period are smaller than those of 
the Newtonian orbital precessions. 

\subsubsection{Comparing with the PN simulations}
\begin{figure*}
\center
\includegraphics[scale=0.70]{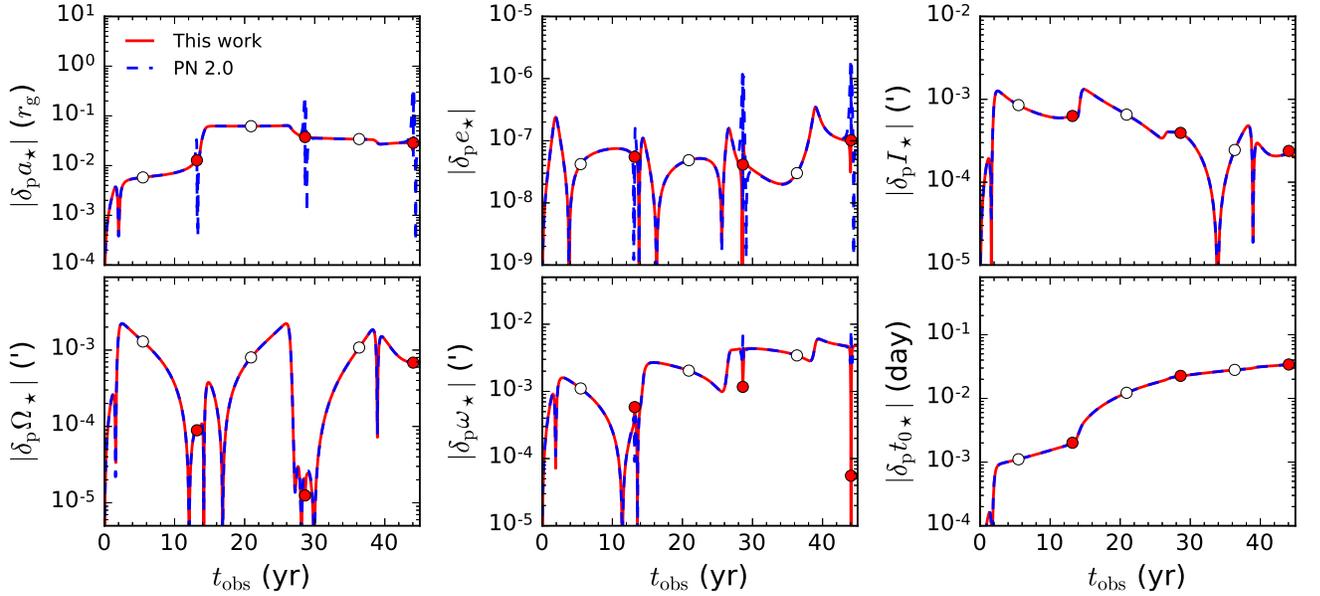}
\caption{Perturbations on the six orbital elements of S2/S0-2 due to the Newtonian attractions of 
S0-102, obtained by post-Newtonian numerical simulations with 2.0 order corrections (blue dashed 
lines) or the method presented in this work (red solid lines). The mass of S0-102 is assumed to be 
$m_\p=0.5\msun$. Red solid and white open circles mark the periapsis and apoapsis passage points of 
S2/S0-2, respectively.
}
\label{fig:ele_S2_com}
\end{figure*}
~\label{subsubsec:com_pn}

The orbital elements of the target star are defined according to its simultaneous position and velocity. Thus, the signals of the Newtonian
perturbations on them may show different behaviours if the adopted spacetime metric is different. It would be interesting to compare the Newtonian perturbations on the orbital elements resulting from our simulations with those from the post-Newtonian (PN) numerical simulations. 
Here, with the same initial conditions, we implement the PN numerical simulations to calculate the orbital evolutions of S2/S0-2 due to the 
gravitational attractions of S0-102. We adopt the PN formalism in~\citet{Kidder95} up to 2.0 order and ignore the spin-related PN terms.
For both the simulations adopting PN formalism and the method in this work, we extract the effects of the Newtonian attraction of S0-102 
from the total effects by examining the differences of the orbital elements when we turn the Newtonian gravity of S0-102 
on and then off (See the extraction method described in Section~\ref{sec:pb_target}).

The comparisons between results of the PN simulation and 
those of this work are showed in Figure~\ref{fig:ele_S2_com}. We can see that these two 
methods predict almost the same evolutions of the perturbations, especially for some of the orbital 
elements, i.e., $I_\star$, $\Omega_\star$ and $t_{0\star}$. For the other elements defined in the orbital plane, i.e. 
$a_\star$, $e_\star$ and $\omega_\star$, there is a relatively large 
discrepancy near each pericenter passage. Such differences can be explained by the fact that the spacetime metric in the 
PN simulations is based on the weak field approximation, while here we use the pure Kerr metric.
Nonetheless, we find that they lead to only a negligible difference in the overall effects of the Newtonian attractions of 
S0-102. For example, after three orbital evolution, the relative difference of the Newtonian 
perturbations on the orbital elements obtained from this work ($\delta_\p \kappa_{\rm hm}$) and 
those from the PN simulations ($\delta_\p \kappa_{\rm pn}$) is about $|\frac{\delta_\p \kappa_{\rm 
hm}-\delta_\p \kappa_{\rm pn}}{\delta_\p \kappa_{\rm pn}}|\lesssim 10^{-2}$, here $\kappa$ is any 
orbital elements of S2/S0-2. The PN simulations result in $\delta_\p f\simeq-4.2'$ at the third 
pericenter. By Equation~\ref{eq:r} and~\ref{eq:z} we find that the resulting perturbations are given 
by $|\delta_\p\mathbf{R}|_{f_\star}\simeq17.5\mu$as and $|\delta_\p Z|_{f_\star}\simeq3.1 \kms$, 
which are consistent with those from our method. However, the PN simulations consider only the 
orbital variations in the local frame of the star. As we additionally include the light tracing 
technique, our simulation has the power of predicting the stellar perturbations on the direct 
observables of the target star, which are also the main focus of this study. 

\subsection{The perturbations on a hypothetical S-star}
~\label{subsec:inner_Sstar}
\begin{figure}
\center
\includegraphics[scale=0.70]{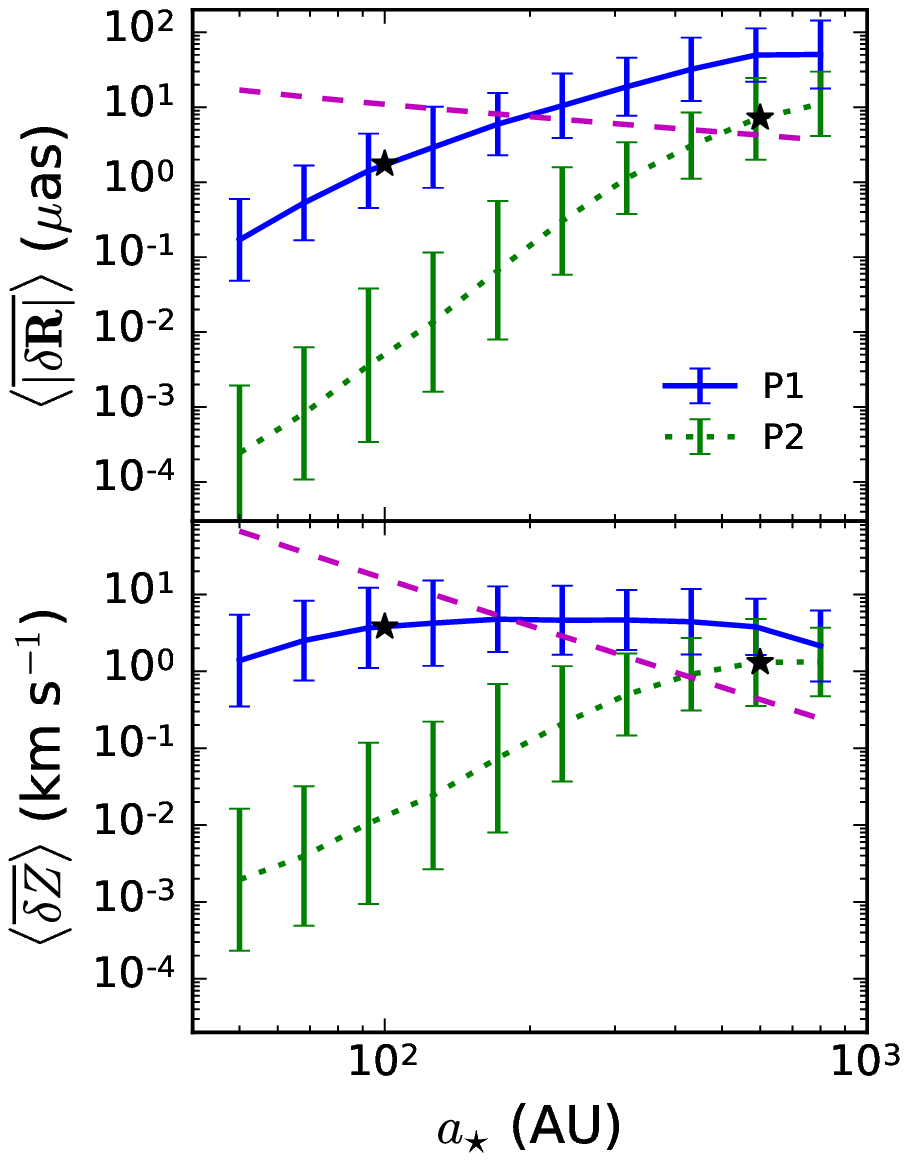}
\caption{Top panel: Perturbations on the apparent position of the target star T1 (with 
$e_\star=0.88$, see Table~\ref{tab:t1}) as a function of $a_\star$. Here 
$\langle \overline{|\delta \mathbf{R}|}\rangle$ is the log-average value of the RMS position 
displacement of the target star in $100$ MC simulations (See 
Equation~\ref{eq:avgrms}). In each MC simulation, the target star is perturbed by a single perturber 
with randomly selected values of $I_\p$, $\Omega_\p$, $\omega_\p$ and $t_{0\p}$. The blue solid and 
green dotted lines show the simulation results of $\langle \overline{|\delta_\p 
\mathbf{R}|}\rangle$ when the perturber is with $a_\p=100\AU$ (the perturber P1) and $a_\p=600\AU$ 
(the perturber P2), respectively. The associated error bars are the standard deviations. The initial 
orbital elements of target stars and perturbers are showed in Table~\ref{tab:t1}. The magenta dashed 
line shows the RMS spin-induced position displacement ($\overline{|\delta_\s \mathbf{R}|}$) in the 
case that $a=0.99$, $i=45^\circ$, and $\epsilon=180^\circ$. The black star symbol in each line marks 
the location where $a_\star=a_\p$. Bottom panel: Similar to the top panel, but for the stellar 
perturbations on the redshift of the target star ($\langle\overline{\delta Z}\rangle$).
}

\label{fig:sg_a_T1}
\end{figure}
\begin{figure}
\center
\includegraphics[scale=0.70]{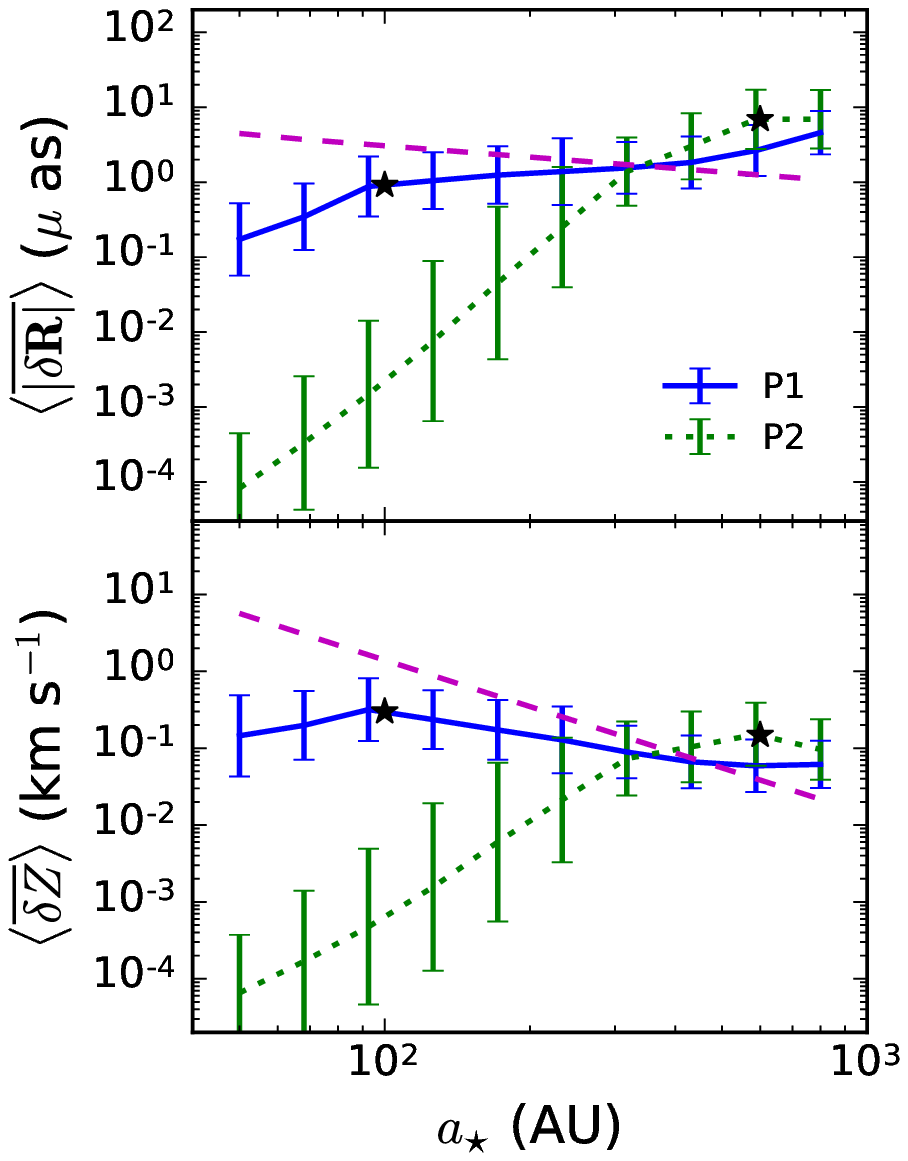}
\caption{Legends similar to those for the Figure~\ref{fig:sg_a_T1}, but for the target star T2 
(with $e_\star=0.3$, see Table~\ref{tab:t1}).
}
\label{fig:sg_a_T2}
\end{figure}

It is plausible that some currently undetected S-stars are located inside the orbit of  
S2/S0-2 or S0-102 in the GC, which may be revealed by the future 
facilities~\citep[e.g.,][]{Zhang13}. These S-stars are better GR probes than S2/S0-2 and may 
be able to put tight constraints on the spin parameters~(e.g.,~\citealt{Angelil10a}; ZLY15).  
Similarly to S2/S0-2, these S-stars are likely to be disturbed by other surrounding stars or 
stellar remnants, and such perturbations should be carefully handled for the unbiased measurements 
of the spin-induced effects. By performing a large number of Monte Carlo (MC) simulations, here we 
study the case that a hypothetical S-star located inside the orbits of S2/S0-2 or S0-102 
($\lesssim800\AU$) is perturbed by a single perturber. As these stars are currently undetected, for 
both the target star and the perturber we consider various initial conditions of them. The details 
of the simulations and the results are showed in the following sections.

\subsubsection{The Monte Carlo Simulations}
We explore the problem by performing the MC numerical simulations as follows: We consider 
a target star with $a_\star$ takes a value between $50\AU$ to $800\AU$, $e_\star=0.88$ (The star 
T1) or $e_\star=0.3$ (The star T2). For each target star, we perform $N_{\rm MC}=100$ MC 
simulations in which it is perturbed by a perturber with randomly selected values of $I_\p$, 
$\Omega_\p$, $\omega_\p$ and $t_{0\p}$\footnote{ We find that the Newtonian 
perturbations depend complexly on the the parameter $I_\p$, $\Omega_\p$, $\omega_\p$ and $t_{0\p}$ 
of the perturber. Here we randomize them as we do not have particular interests in the details of 
the dependence on these parameters but focus only on their average effects of the Newtonian 
perturbations.} over three orbits. The orbital semimajor axis of the perturber 
in these MC simulations are $a_\p=100\AU$ (The perturber P1) or $600\AU$ (The perturber P2). The 
initial conditions of the target stars and the perturbers are listed in Table~\ref{tab:t1}. We 
estimate the log-average value of the RMS position displacement in these MC 
simulations, i.e., $\langle\overline{|\delta \mathbf{R}|}\rangle$, by 
\be
\ba
\log \langle\overline{|\delta \mathbf{R}|}\rangle=\sum_{j}^{N_{\rm MC}}\frac{\log\overline{ 
|\delta \mathbf{R}|}_j}{N_{\rm MC}},
\label{eq:avgrms}
\ea
\ee
here $N_{\rm MC}=100$, $|\delta \mathbf{R}|_j$ is the RMS perturbations on the position signal in 
the $j$-th MC simulation. Similarly, we can define the log-average value for the redshift signal, 
i.e., $\langle \overline{\delta Z}\rangle$, or any other quantities of the target star. The 
simulation results of $\langle \overline{|\delta_\p \mathbf{R}|}\rangle$ and 
$\langle\overline{\delta_\p Z}\rangle$ for the target star T1 and T2 are showed in 
Figure~\ref{fig:sg_a_T1} and~\ref{fig:sg_a_T2}, respectively. The associated error bars show the 
standard deviations (about $1-1.5\,$dex) due to the randomly selected initial values 
of $\Omega_\p,\Upsilon_\p,I_\p$ and $t_{0\p}$. Note that for the spin-induced 
signals, we simply have $\langle\overline{|\delta_\s \mathbf{R}|}\rangle=\overline{|\delta_\s 
\mathbf{R}|}$ and $\langle \overline{\delta_\s Z}\rangle=\overline{\delta_\s Z}$.

\subsubsection{Results}
~\label{subsubsec:hs_results}

From Figure~\ref{fig:sg_a_T1} and~\ref{fig:sg_a_T2}, it appears that the scaling 
of $\langle\overline{|\delta_\p \mathbf{R}|}\rangle$ with $a_\star$ is different from those of 
$\langle \overline{\delta_\p Z}\rangle$. Such difference can be explained as follows. According to 
Section~\ref{subsubsec:pb_orb_S2}, the RMS Newtonian perturbations on the observables are related to 
the perturbations on the true anomaly of the target star by $\overline{|\delta_\p 
\mathbf{R}|}\propto a_\star \overline{\delta_\p f_\star}$ 
and $\overline{\delta_\p Z}\propto a_\star^{-1/2} \overline{\delta_\p f_\star}$, (See 
Equation~\ref{eq:r}, and \ref{eq:z}). After averaging the different MC runs, we have 
$\langle\overline{|\delta_\p \mathbf{R}|}\rangle \propto a_\star\langle\overline{\delta_\p 
f_\star}\rangle$ and $\langle \overline{\delta_\p Z}\rangle\propto 
a_\star^{-1/2}\langle\overline{\delta_\p f_\star}\rangle$. Here $\langle \overline{\delta_\p 
f_\star}\rangle$  is the log-average value estimated by method similar to 
Equation~\ref{eq:avgrms}. Thus, $\langle\overline{|\delta_\p \mathbf{R}|}\rangle\propto 
a_\star^{3/2}\langle \overline{\delta_\p Z}\rangle$, i.e., the scaling relations of these two 
signals are different by a factor of $a_\star^{3/2}$. 

The Newtonian perturbations depend on the relative location of the perturber from the target star. 
From Figure~\ref{fig:sg_a_T1} and~\ref{fig:sg_a_T2}, a perturber impose stronger Newtonian 
perturbations on the observables of a target star located around or outside of its orbit 
($a_\star\ga a_\p$), than that located inside of its orbit ($a_\star\la \lambda 
a_\p$, $\lambda=0.5\sim0.8$ is a factor determined by the simulations). For target stars located 
inside the orbit of the perturber, the Newtonian perturbations are increasing functions of 
$a_\star$. For example, in Figure~\ref{fig:sg_a_T1}, for the perturber with $a_\p=600\AU$ (the green 
dotted lines), $\langle\overline{|\delta_\p \mathbf{R}|}\rangle$ increases from 
$2.5\times10^{-4}\mu$as to $3.1\mu$as, and $\langle\overline{|\delta_\p Z|}\rangle$ 
increases from $0.002\kms$ to $0.9\kms$, if the $a_\star$ of the target star increases 
from $50\AU$ to $432\AU$. 

The Newtonian perturbations depend on other parameters of the target star and the 
perturber, which we describe them briefly here: (1) Both the Newtonian perturbation 
and the spin-induced effects strongly depend on the eccentricity of the target star. From 
Figure~\ref{fig:sg_a_T1} and~\ref{fig:sg_a_T2}, we can see that the target stars with 
$e_\star=0.3$ felt much smaller spin-induced effects and Newtonian perturbations than those 
target stars with $e_\star=0.88$. (2) The Newtonian perturbations are proportional to the mass of 
the perturber, i.e., $\langle\overline{|\delta_\p \mathbf{R}|}\rangle\propto m_\p$ and 
$\langle\overline{|\delta_\p Z|}\rangle\propto m_\p$ (See also Section~\ref{subsec:S2} or 
Figure~\ref{fig:mass_S0-102}). We find that such relation remain true if the mass of the perturber 
is in the range of $0.1\msun<m_\p<100\msun$. (3) The average values of stellar perturbations depend 
weakly on the eccentricity of the perturber, i.e., $e_\p$. For example, if the 
eccentricities of the perturbers in the MC simulations of Figure~\ref{fig:sg_a_T1} 
and~\ref{fig:sg_a_T2} are replaced by $e_\p=0.3$, we find 
that the results are quite similar. (4) The Newtonian perturbations depend complexly on other 
parameters related to the orbital configurations of the target star, i.e., $I_\star$, 
$\Omega_\star$, $\omega_\star$ and $f_\star$. We found that the Newtonian perturbations may differ 
by a factor of several to one order of magnitude if adopting different values of these parameters.

We can see that the Newtonian perturbations by a $10\msun$ perturber is already large enough to 
obscure the spin-induced signals. For the simulations showed in Figure~\ref{fig:sg_a_T1} 
and~\ref{fig:sg_a_T2}, the spin-induced effects on the apparent position (or the redshift) can be 
drowned by the Newtonian perturbations if the orbital semimajor axis of the target star is larger 
than $260\sim500\AU$ (or $200\sim430\AU$). 

\section{The Newtonian perturbations of a star cluster}
~\label{sec:nw_cluster}
\begin{figure}
\center
\includegraphics[scale=0.70]{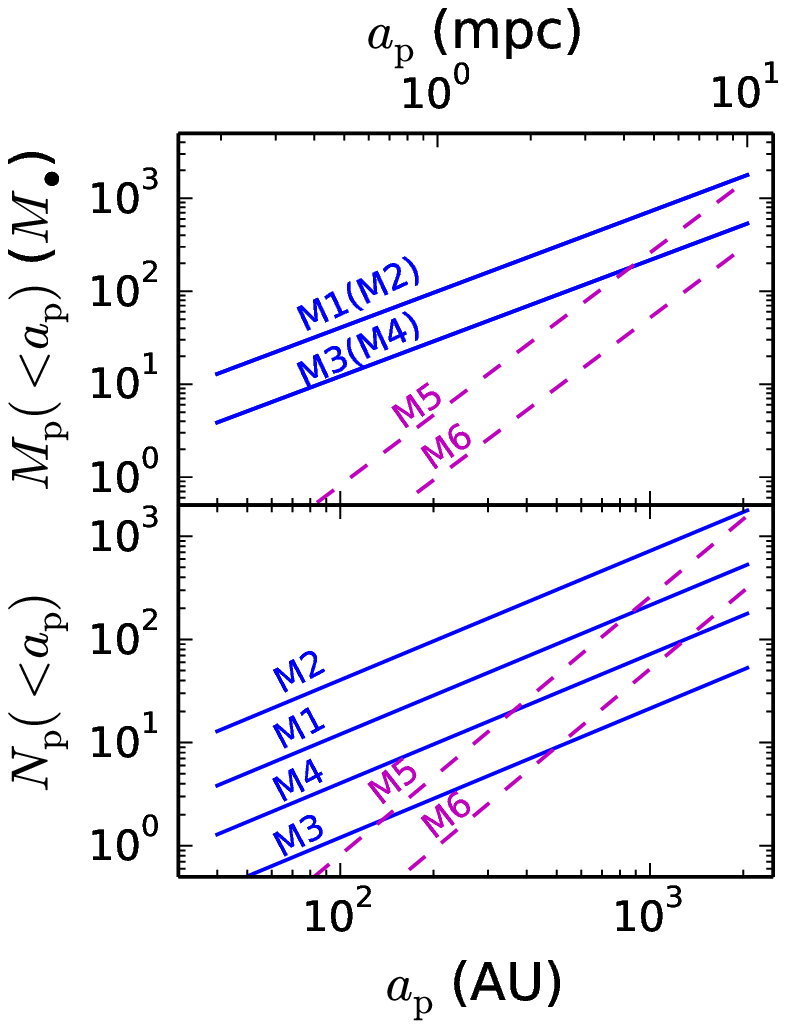}
\caption{Total mass [$M_\p(<a_\p)$, top panel] and number [$N_\p(<a_\p)$, bottom panel]
 of perturbers with semimajor axis smaller than $a_\p$ for clusters in different models. The 
blue solid and the magenta dashed lines correspond to the models with the Bahcall-Wolf profiles 
($\gamma=7/4$, model M1-M4) and the core-like profiles ($\gamma=0.5$, model M5-M6), respectively.
}
\label{fig:profile}
\end{figure}

\begin{table}
\caption{Parameters of the Clusters in Different Models.}
\centering
\begin{tabular}{cccccccccccccccccc}
\hline
Model  & $M_{\p}$ & $M_{1}$ & $\gamma$  &
$\beta$  & $m_\p$ & $N_\p$ & $N_{\rm MC}$ \\
&[1]&[2]&[3]& [4]& [5]  &[6] &[7] \\
\hline
M1   & 1780   & 100 & 1.75  & 0.5   & 10   & 178    & 80  \\
M2   & 1780   & 100 & 1.75 & 0.5    &  1   & 1780   & 8   \\
M3   & 530     & 30  & 1.75 & 0.5   &  10   & 53    & 280\\
M4  & 530     & 30   & 1.75 & 0.5   &  1   & 530    & 28 \\
M5    & 1581    & 5  & 0.5  & -0.5   &  1  & 1581    & 10 \\
M6   & 316    &  1   & 0.5   & -0.5   &  1  &  316 &  48 \\ 
\end{tabular}
\label{tab:t3}
\tablecomments{The initial conditions of the clusters in different models. Col.[1]: The 
total mass of the cluster in units of $\msun$. Note that the perturbers in the clusters are with 
$40{\AU}<a_{\p}<2062\AU$ ($0.2$ mpc $\la a_\p \la 10$ mpc); Col.[2]: Total mass of the perturbers 
with $0.2$ mpc $\la a_\p \la 1$ mpc, i.e., $M_1=M_\p(<1$ mpc$)$, in units of $\msun$; Col.[3]: 
Slope of the density profile, i.e., $n(r)\propto r^{-\gamma}$; Col.[4]: The velocity anisotropy, 
given by $\beta=1-\sigma_{\rm tr}^2/\sigma_{\rm los}^2$, here $\sigma_{\rm tr}$ and $\sigma_{\rm 
los}$ is the velocity dispersion in the transverse and the line of sight direction, respectively; 
Col.[5]: The mass of the perturber in units of $\msun$; Col.[6]: The total number of the perturbers 
in the cluster, i.e., $N_\p=M_\p/m_\p$; Col.[7]: The total number of MC simulations. 
}
\end{table}

\begin{table}
\caption{Critical Values of the Target Stars in Different Models.}
\centering
\begin{tabular}{cccccccccc}\hline
\multirow{3}{1.0cm}{Model}  &
\multicolumn{4}{c}{$e_\star=0.88$} &&
\multicolumn{4}{c}{$e_\star=0.3$}\\
\cline{2-5}\cline{7-10}
& $a_{\star}^{\rc}$ &$r_{{\rm per},\star}^{\rc}$ & $a_{\star}^{\zc}$ & $r_{{\rm per},\star}^{\zc}$  &
& $a_{\star}^{\rc}$ &$r_{{\rm per},\star}^{\rc}$ & $a_{\star}^{\zc}$ & $r_{{\rm per},\star}^{\zc}$  \\
&[1]&[2]&[3]& [4]& & [5]  &[6] &[7]  & [8]  \\
\hline
M1    & 120 &360 & 100 & 300 && 100 & 1770 & 110 &1950\\
M2   & 170  &520 & 130 & 400 && 110 &1950 & 120 & 2130 \\
M3   & 170  &520  & 130 & 400 && 130 &2310 & 150  & 2660\\
M4   & 230 &700 & 170 & 520 && 170 &3020 & 190 &3370 \\
M5   &270 &820 & 200 & 610 && 200 & 3550& 210 & 3730\\
M6   &390 &1190 & 330 &1000 && 300 &5320 & 350 & 6210\\ 
\end{tabular}
\label{tab:t4}
\tablecomments{The critical orbital semimajor axis or the distance at the pericenter of the target stars where the effects 
of Newtonian attraction equal to the spin-induced ones. Col.[1-4]: 
$a^{\rm rc}_{\star}$ (or $a^{\rm zc}_\star$) is the orbital semimajor axis when 
$\langle\overline{|\delta_\p\mathbf{R}|}\rangle\simeq\overline{|\delta_\s 
\mathbf{R}|}$ (or $\langle\overline{|\delta_\p Z|}\rangle\simeq\overline{|\delta_\s Z|}$) for the 
target star T1 (with $e_\star=0.88$), in unit of $\AU$. $r^{\rm rc}_{{\rm per},\star}=a^{\rm rc}_{\star}(1-e_\star)$ 
[or $r^{\rm zc}_{{\rm per},\star}=a^{\rm zc}_\star(1-e_\star)$]
is the corresponding distance of the pericenter, in unit of $r_\g\sim0.04\AU$.
Col.[5-8]: Similar to Col.[1-4] but for the target star T2 (with $e_\star=0.3$). For the parameters of these target stars 
see Table~\ref{tab:t1}. 
}
\end{table}

Currently, the mass distribution within milli-parsec 
scale in the GC along with its actual composition remain largely uncertain. The 
infrared imaging and spectroscopic observations in the past two decades have revealed over 
thousands of brightest stars in the inner parsec distance~\citep[e.g.,][]{Genzel10,Schodel09}. Most 
of these observed stars can be classified into two distinctive categories: (1) The early-type 
stars, which are the young ($\sim10$Myr), massive ($\gtrsim 7\msun$) and main-sequence O-type or 
B-type stars. Although these stars are rare ~\citep[numbers of about several 
hundred, e.g.,][]{Paumard06}, they dominate the total luminosity within the inner parsec of the GC. 
(2) The late-type stars, which are old (several Gyrs), K-type or 
M-type giant stars with masses $1\sim2\msun$. These stars dominate the total star counts 
observed in the inner parsec of the GC. Due to their long lifetime, it is believed that they are 
the most promising tracers of the stellar distributions in the GC.

The proper motion observations of the late-type stars suggest that the extended mass within one 
parsec from the GC is given by $0.5\sim1.5\times10^6\msun$~\citep[e.g.,][]{Schodel09}. 
However, the radial distributions of the extended mass are not well constrained.
Theoretical works expect that if the stars around the MBH are dynamically relaxed by two body 
interactions, cusp profiles should appear, i.e., $n(r)\propto r^{-\gamma}$, with 
$\gamma=3/2\sim7/4$~\citep{BW76,BW77}. However, recent observations of the late-type stars 
suggest a flattened core-like profiles with $\gamma=0\sim1$ towards the inner 
region~\citep[e.g.,][]{Do09,Buchholz09,Bartko10,Do13a, Do13b}. The deficit of the inner stars is 
currently not well understood. It may suggest that stars/stellar remnants in the inner parsec are 
not in the equilibrium state, or a significant number of the late-type stars in this region are 
destroyed by stellar collisions~\citep[e.g.,][]{Alexander99,Freitag06,Dale09}.

It is expected that a cluster of stellar mass black holes (with masses of $\sim 10\msun$) may 
exist in the vicinity of the MBH, if the two-body relaxation time is less than a few Gyr in the 
GC~\citep{Hopman06,Alexander09}. The stellar mass black holes may form by the collapses and 
explosions of the early-type stars at the end of their main-sequence lives, which later concentrate 
towards the center through mass segregation~\citep[e.g.,][]{Freitag06}. However, so far it remains 
largely unclear whether they dominate the mill-parsec scales.

Due to the large uncertainties in this region, we explore the stellar perturbations on the 
observables of a target star surrounded by a star cluster with some possible mass distributions and 
its composition. The details of the simulations and the results are showed in the following 
sections.

\subsection{The model parameters}
\begin{figure*}
\center
\includegraphics[scale=0.70]{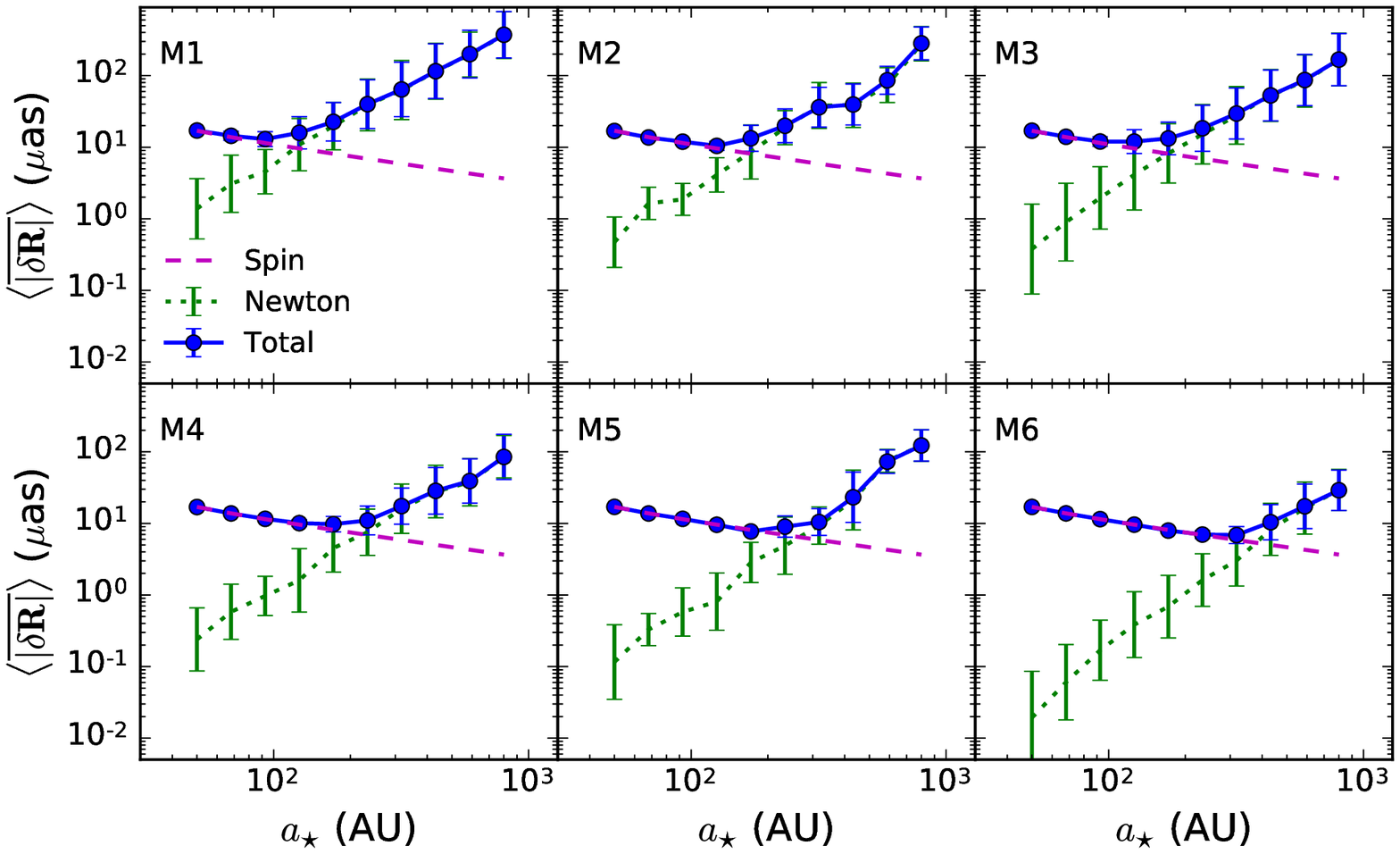}
\caption{Perturbations on the apparent position of target star T1 (see Table~\ref{tab:t1}) as 
a function of $a_\star$ in different model clusters. $\langle\overline{|\delta_\p\mathbf{ 
R}|}\rangle$ is defined by Equation~\ref{eq:avgrms},  i.e., the log-average value of the RMS position displacement in different 
MC simulations. In each panels, the magenta dashed, 
green dotted, and the blue solid lines show the spin-induced perturbations, stellar perturbations, 
and the total perturbations, respectively.}
\label{fig:sg_cluster_rms}
\end{figure*}
\begin{figure*}
\center
\includegraphics[scale=0.70]{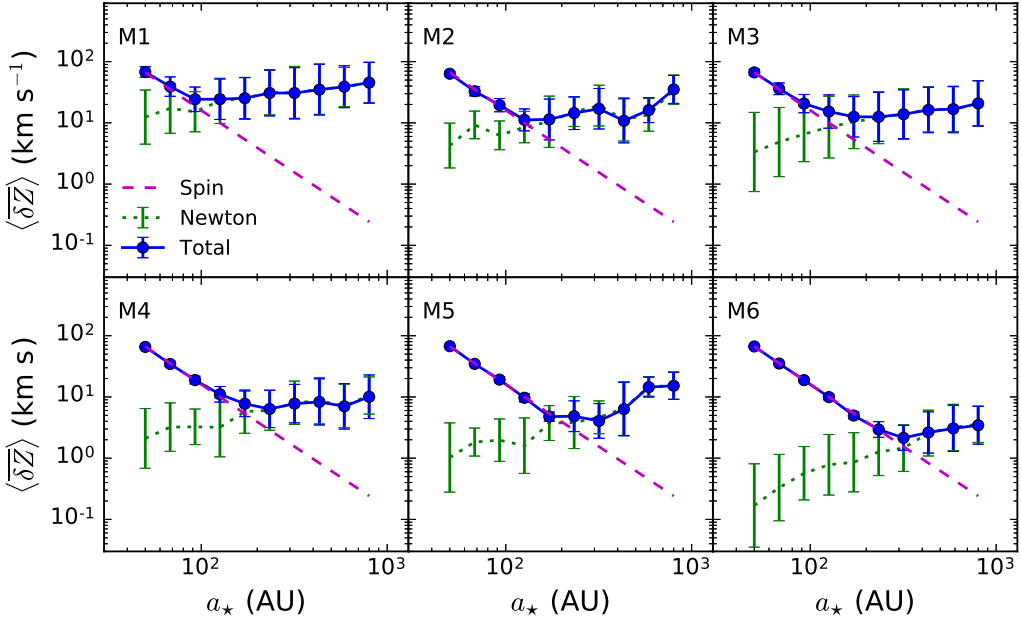}
\caption{Legends similar to those for the Figure~\ref{fig:sg_cluster_rms}, but for the 
perturbations on redshift signal of the target star T1.
}
\label{fig:sg_cluster_zms}
\end{figure*}
\begin{figure*}
\center
\includegraphics[scale=0.65]{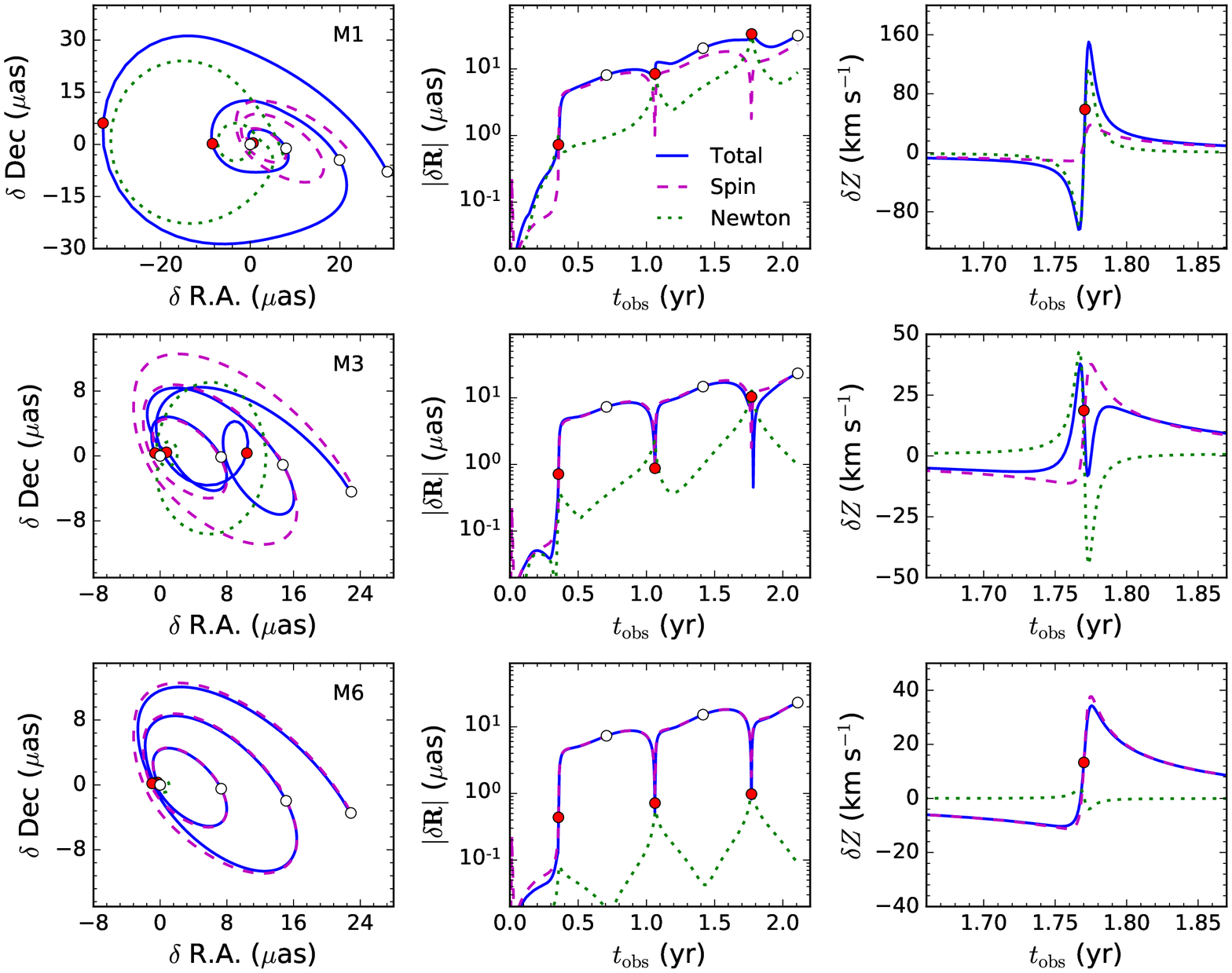}
\caption{Perturbations on the apparent position [$\delta \mathbf{R}=(\delta$R.A.,
$\delta$Dec$)$, left panels], its distance ($|\delta \mathbf{R}|$, middle panels) and the redshift 
($\delta Z$, right panels) of an example target star T1 (see Table~\ref{tab:t1}) with 
$a_\star=126\AU$ over three orbits ($\sim2.2$yr). The top, middle and bottom panels show the 
results when the target star is surrounded by the cluster M1, M3 and M6, respectively. In each 
panel, the spin-induced perturbations, Newtonian perturbations and the combinations of the 
above two are showed in the magenta dashed, green dotted, and the blue solid lines, respectively. 
Red solid and white open circles mark the periapsis and apoapsis passage points of the 
target star, respectively. Note that the right panels show the difference in the redshift, i.e.,  
$\delta Z$, near the third pericenter passage of the target star. The target star is selected out 
as 
a typical star from a number of $N_{\rm MC}$ MC runs 
with $\overline{|\delta_\p\mathbf{R}|}\sim\langle\overline{|\delta_\p\mathbf{ R}|}\rangle$ and 
$\overline{\delta_\p Z}\sim\langle\overline{\delta_\p Z}\rangle$, here 
$\langle\overline{|\delta_\p\mathbf{R}|}\rangle$ and $\langle\overline{\delta_\p Z}\rangle$ of the 
target stars are showed in Table~\ref{tab:t5}.
}
\label{fig:sg_cluster}
\end{figure*}

We adopt six models (M1-M6, see Table~\ref{tab:t3}) with different initial conditions of the 
star cluster to cope with the large uncertainties in the mass profile within several mpc from the 
MBH. The details of the model setups are described as follows.

We assume that the cluster consists of perturbers with equal mass $m_\p$, orbital 
semimajor axis and eccentricity, i.e., $a_\p$ and $e_\p$, following the distribution functions 
$g(a_\p)$ and $h(e_\p^2)$, respectively. We assume that the perturbers are with randomly selected 
initial values of the inclination $I_\p$, position angle of ascending node $\Omega_\p$, angle to periapsis $\omega_\p$ 
and the time of pericenter passage 
$t_{0\p}$. We assume that the number density of stars in the cluster is given by $n(r)\propto 
r^{-\gamma}$, and their velocity 
anisotropy is given by $\beta=1-\sigma_{\rm tr}^2/\sigma_{\rm los}^2$, here $\sigma_{\rm tr}$ and 
$\sigma_{\rm los}$ is the velocity dispersion in the transverse and the line of sight direction, 
respectively. Then it turns out~\citep{Merritt10} that $a_{\p}$, $e_{\p}$ of the perturbers in the 
cluster follow a distribution $g(a_{\p})\propto a_{\p}^{2-\gamma}$, and $h(e_{\p}^2)\propto 
(1-e_{\p}^2)^{-\beta}$,  $\beta\le \gamma-1/2$. Then the total mass and number of perturbers with 
semimajor axis smaller than $a_\p$, i.e., $M_\p(<a_\p)$ and $N_\p(<a_\p)$, is given by $M_\p 
(<a_\p)=M_1 [a_\p/ (1~{\rm mpc})]^{3-\gamma}$ and $N_\p(<a_\p)=M_\p(<a_\p)/m_\p$, respectively. 
Here $M_1=M_\p(<{1~{\rm mpc}})$ is the mass of perturbers with $a_\p<1$ mpc. 

We restrict that $40{\AU}<a_\p<2062\AU$ (or $0.2$ mpc $\la a_{\p}\la 10$ mpc), and denote the 
total mass and number of perturbers in the cluster by $M_\p$ and $N_\p$. The outer boundary 
($10$ mpc) are found large enough for the convergence of the simulation results in this work.
The inner boundary ($0.2$ mpc) is set to avoid perturbers too close to the MBH, which are 
found slow down the numerical simulation quite significantly and the remove of them cause 
negligible difference on the simulation results. We avoid those stars of which the periapsis 
distance is within their tidal radius, i.e.,  $\simeq (\eta^2 \bh/m_\p)^{1/3} 
(m_\p/\msun)^{0.47}R_\odot$, where $\eta=2.21$~\citep{Magorrian99}. We also avoid the stars with 
orbital gravitational wave radiation timescale $T_{\rm GW}<100$Myr, as the total number of these 
stars may be substantially suppressed due to the rapid orbital decay. 

It is still unknown the extent to which the stars and stellar remnants around the vicinity of the 
MBH are dynamically relaxed by two body interactions. Thus, we consider two extreme cases: (1) They 
are dynamically relaxed, so that the density profile approaches to the Bahcall-Wolf cusp profile, 
i.e. $\gamma=1.75$ (model M1-M4). In this case, it is possible that the stellar mass black holes 
may dominate the vicinity of the MBH due to the mass segregation 
effects~\citep[e.g.,][]{Hopman06,Alexander09}. We assume that all of the perturbers in the cluster 
are either low mass main-sequence stars with $m_\p=1\msun$ or stellar mass black holes with 
$m_\p=10\msun$. We set $M_{1}=100\msun$ or $30\msun$ in each of these cases. Then the total 
mass of the cluster is given by $M_{\p}=1780\msun$ (model M1 and M2) or $530\msun$ (model 
M3 and M4). (2) The mass distribution in the vicinity of the MBH have not yet reached the 
equilibrium state by two body relaxation; then, the density profile is likely flatter than the BW 
cusp profiles. We assume that the mass distribution follows a core-like profile with $\gamma=0.5$ 
(model M5 and M6). In this case, we assume that the perturbers are all low mass main-sequence stars 
with $m_\p=1\msun$ and $M_{1}=5\msun$ or $M_1=1\msun$. Then the total stellar mass is given by 
$M_{\p}=1581\msun$ (model M5) or $316\msun$ (model M6). The total mass and the number of perturbers 
with orbital semimajor axis smaller than $a_\p$, i.e., $M_\p(<a_\p)$ and 
$N_\p(<a_\p)$, for all the models in Table~\ref{tab:t3} are showed in Figure~\ref{fig:profile}.

As we find that the results in this section are insensitive to the eccentricity distribution of 
the perturber (See also Section~\ref{subsubsec:hs_results}), we assume $\beta=0.5$ for models 
with $\gamma=1.75$ and $\beta=-0.5$ for models with $\gamma=0.5$ in order to fulfill the condition 
$\beta\le \gamma-1/2$. 

A target star is assumed embedded in each of these clusters. Similarly to 
Section~\ref{subsec:inner_Sstar}, we consider the target star T1 or T2 (See 
Table~\ref{tab:t1}). We integrate the perturbations on the target star over three orbits. If 
without otherwise specified, we assume that the spin parameters are given by $a=0.99$, 
$i=45^\circ$, and $\epsilon=180^\circ$. For each model in Table~\ref{tab:t3}, $N_{\rm MC}$ 
independent realizations are performed such that the total number of perturbers in the combined set 
of integration is $\sim15000$. Then we estimate the changes of the orbital elements and 
the observables of the target star due to the Newtonian and spin-induced perturbations. The details 
of the results are described in the following sections.

\begin{table}
\caption{Stellar Perturbations on the Motion and Signals of the Target Star T1 with 
$a_\star=126\AU$.}
\centering
\begin{tabular}{ccccccccc}\hline\\[-6pt]
Name  &  
$\langle\overline{|\delta_\p f_\star|}\rangle$ &  
$\langle\overline{|\delta_\p I_\star|}\rangle$ & 
$\langle\overline{|\delta_\p \Omega_\star|}\rangle$ & 	
$\langle\overline{|\delta_\p \omega_\star|}\rangle$ &
$\langle\overline{|\delta_\p\mathbf{R}|}\rangle$ & 
$\langle\overline{|\delta_\p Z|}\rangle$\\
& [1] & [2]& [3]& [4]& [5]& [6] \\[2pt]
\hline
M1   & 14.7 & 0.23 & 0.14 & 0.23 & 10.9 & 22.8  \\
M2   & 5.5  & 0.08 & 0.06 & 0.09 & 4.1  & 8.5  \\
M3   & 5.5  & 0.1  & 0.06 & 0.1  &4.1 & 8.5   \\
M4   & 2.1  & 0.04 & 0.03 & 0.05 &1.6 & 3.2   \\
M5   & 1.0  & 0.03 & 0.01 &0.02  &0.8 & 1.6   \\
M6   & 0.5  & 0.01 & 0.004 &0.008 &0.4 & 0.8   \\ 
\end{tabular}
\label{tab:t5}
\tablecomments{The stellar perturbations on the orbital motion and observables of a target star T1 with $a_\star=126\AU$.
The values for the orbital elements of the target star ($f_\star$, $I_\star$, 
$\Omega_\star$ and $\omega_\star$) are 
showed in Col.[1-4] in units of ($'$). The values for the position and redshift signals of the 
target star are showed in Col.[5] (in units of $\mu$as) and Col.[6] (in units of $\kms$), 
respectively. For each values in the table, the standard deviations is $\sim0.5-1$dex. 
As a comparison, the RMS spin position and redshift signals of this target star are 
given by $\overline{|\delta_\s \mathbf{R}|}=9.6\mu$as and $\overline{|\delta_\s Z|}=10\kms$, 
respectively. 
}
\end{table}

\subsection{The perturbed observables of the target star}
~\label{subsec:cluster_pb_ob}
We denote $\langle \overline{|\delta \mathbf{R}|}\rangle$ and $\langle \overline{\delta Z}\rangle$ 
as the log-average values of the perturbations of the clusters (Similar to 
Equation~\ref{eq:avgrms}), for the position and redshift signals, respectively. 
The simulation results of $\langle \overline{|\delta \mathbf{R}|}\rangle$ and $\langle 
\overline{\delta Z}\rangle$ of the target star T1 in clusters M1-M6 are showed 
in Figure~\ref{fig:sg_cluster_rms} and~\ref{fig:sg_cluster_zms}, respectively. We find that 
similar results can be obtained for the target star T2. Note that both the Newtonian and the 
spin-induced perturbations felt by the target star T2 are smaller than those by T1, simply due to 
its smaller eccentricity ($e_\star=0.3$).

According to Figure~\ref{fig:sg_cluster_rms} and~\ref{fig:sg_cluster_zms}, both $\langle 
\overline{|\delta_\p \mathbf{R}|}\rangle$ and $\langle \overline{\delta_\p Z}\rangle$ are 
increasing functions of $a_\star$, although the slope index of $\langle \overline{|\delta_\p 
\mathbf{R}|}\rangle$ is larger than that of $\langle \overline{\delta_\p Z}\rangle$ as
$\langle \overline{|\delta_\p \mathbf{R}|}\rangle/\langle \overline{\delta_\p Z}\rangle\propto 
a_\star^{3/2}$ (See also Section~\ref{subsubsec:hs_results}). As the spin-induced 
effects on both position and redshift are decreasing functions of $a_\star$, i.e., 
$\overline{|\delta_\s\mathbf{R}|}\propto a_\star^{-3/2}$ and $\overline{\delta_\s Z}\propto 
a_\star^{-2}$, for each model there is a critical orbital semimajor axis of the target star
that the effects of Newtonian attraction and the spin equal to each other (See more details in 
Section~\ref{subsec:cluster_crit}). The combined effects of them (See the blue solid line in each 
panel of Figure~\ref{fig:sg_cluster_rms} and~\ref{fig:sg_cluster_zms}) are thus dominated by the 
spin-induced effects in the inner region and the Newtonian perturbations in the outer region.

We find that, for a given total mass $M_\p$, the Newtonian perturbations of a cluster of low mass 
stars are smaller than that of stellar mass black holes~\citep[see also][]{Merritt10}. 
For example, the stellar perturbations of the cluster M1 (with $m_\p=10\msun$ and $N_\p=178$) are 
larger than those of the cluster M2 (with $m_\p=1\msun$ and $N_\p=1780$) by a factor of $\sim2$. 
The reason is probably that M1 consists by more perturbers than M3, such that its potential is more 
isotropic and the induced stellar perturbations are less significant. Similar results can be 
obtained if comparing model M3 with M4.

As showed in Section~\ref{sec:nw_single}, the Newtonian perturbations felt by a target star are 
mainly attributed to perturbers located around or inside of its orbit ($a_\p\la a_\star$). 
For a given total mass $M_\p$, stellar perturbations of the clusters with cusp profile are 
larger than those with core profile, because the former contains more perturbers in the inner 
region than the latter. For example, the mass of cluster M2 is quite similar to that of M5, but the 
stellar perturbations of the cluster M2 is about $4-5$ times larger than those of the cluster M5 
(See Figure~\ref{fig:sg_cluster_rms},~\ref{fig:sg_cluster_zms} and Table~\ref{tab:t5}). 

The observables of the target stars with $a_\star\sim100-400\AU$ may be either dominated by 
Newtonian or spin-induced perturbations, depending on the details of the cluster. 
Figure~\ref{fig:sg_cluster} shows the perturbations on the observables of a target star T1 with 
$a_\star=126\AU$ for three models of the cluster (M1, M2 and M3). The target star in each model is 
selected out from $N_{\rm MC}$ MC runs, with 
$\overline{|\delta_\p\mathbf{R}|}\sim\langle\overline{|\delta_\p\mathbf{R}|}\rangle$ and 
$\overline{\delta_\p Z}\sim\langle\overline{\delta_\p Z}\rangle$ (These values are showed 
in Table~\ref{tab:t5}). From the top to bottom panels of Figure~\ref{fig:sg_cluster}, 
the dominating factor of the signal gradually changes from stellar to spin-induced perturbations. 
Figure~\ref{fig:sg_cluster} suggests that for a target star in a cluster: (1) The stellar 
perturbations on the apparent position and the redshift always peak around the periapsis passage, 
which are mainly caused by the perturbed orbital period (See Section~\ref{subsubsec:pb_orb_S2} 
and the following Section); (2) The combined perturbations are not similar to
either Newtonian or spin-induced ones if they are comparable to each other, i.e., 
$\overline{|\delta_\p \mathbf{R}|}\sim\overline{|\delta_\s \mathbf{R}|}$ or $\overline{\delta_\p 
Z}\sim\overline{\delta_\s Z}$. Thus, a complex morphology of the signal strongly suggests that the 
contaminations by the stellar perturbations occur. (3) The details of the morphologies/evolutions 
of the Newtonian perturbations are quite different with respect to the spin-induced effects. 
According to their distinctive features, in principle, they can be separated from each other.

\subsection{The perturbed orbital elements of the target star}
We describe the results of the stellar perturbations on the orbital elements in this 
section. We find that $\langle\overline{\delta_\p \kappa}\rangle\propto a_\star^\epsilon$, where 
$\kappa$ is any orbital elements and $\epsilon$ is a slope index depending on the details of the 
cluster. For models in Table~\ref{tab:t3}, we found 
$\epsilon\sim1.9-2.5$ if $\kappa=a_\star$, and $\epsilon\sim0.7-1.5$ if $\kappa=e_\star$, 
$I_\star$, $\Omega_\star$, $\omega_\star$ or $f_\star$. The dependence of 
$\langle\overline{\delta_\p \kappa}\rangle$ on the parameters of the clusters are quite similar to 
those of the observables (See Table~\ref{tab:t5}).

Table~\ref{tab:t5} shows the log-average stellar perturbations on the orbital elements and 
observables of the target star T1 with $a_\star=126\AU$. From Table~\ref{tab:t5}, the 
stellar perturbations cause the precessions of both the argument of periapsis ($\langle 
\overline{\delta_\p\omega_\star|}\rangle$) and the orientation of the orbital plane (described by 
$\langle \overline{\delta_\p 
I_\star|}\rangle$ and $\langle \overline{\delta_\p\Omega_\star|}\rangle$).
We found that these results are roughly consistent with the following analytical arguments. The 
Newtonian precession of the argument of pericenter in each revolution is given by~\citep{Madigan11} 
\be\ba
 \delta_\p\omega_\star
=1.72'\times\frac{\mathscr{F}(e_\star,\gamma)}{2-\gamma}\frac{4\times10^6\msun}{\bh}\times\frac
{ M_\p(r<a_\star)} {10^3\msun}\\
~\label{eq:dpomstar}
\ea
\ee
Here $\mathscr{F}(e_\star,\gamma)$ is a factor depending on the eccentricity of the target 
star and the density profile. In the case $\gamma=1.75$, it is given by~\citep{Madigan11}
\be\ba
\mathscr{F}^{-1}(e_\star,1.75)=0.681+\frac{0.975}{\sqrt{1-e_\star}}+0.373(1-e_\star)
\ea
\ee
$M_\p(r<a_\star)$ is the enclosed mass within a radii of $r=a_\star$. For a target star with 
$a_\star=126\AU$ in cluster M1 or M2, $M_\p(r<126\AU)\sim40\msun$\footnote{Note that in cluster M1 
or M2, the total mass of perturbers with orbital semimajor axis less than $126\AU$ is  
$M_\p(a_\p<126\AU)\sim54\msun$. The $M_\p(r<a_\star)$ is somewhat smaller 
than $M_\p(a_\p<a_\star)$, as at any given moment it is more likely to find perturbers near the 
apocenter of their eccentric orbits.}. After three orbits, its RMS Newtonian precession is 
$\overline{\delta_\p\omega_\star} \simeq 3\times \frac{2}{3}\delta_\p\omega_\star\simeq 0.16'$, 
which is roughly consistent with the numerical results in model M1 or M2. Note that 
Equation~\ref{eq:dpomstar} assume that the potential of the cluster is smooth and isotropic, which 
may not be fully satisfied in the simulation. The simulated clusters consist by a finite 
number of perturbers and thus the potential is somewhat anisotropy. This may explain the 
discrepancies between the results of simulations and that predicted by Equation~\ref{eq:dpomstar}. 

The anisotropy of the potential can also lead to effects of the resonant relaxations, which cause  
the precession of the orbital plane of the target star~\citep[e.g.,][]{RT96,Hopman06}. 
According to ~\citealt{Merritt10} (its Equation 26 and 27), in each revolution,
\be
\frac{|\delta_\p \mathbf{L}|}{L_c}\simeq \beta_v \frac{m_\p}{\bh}\sqrt{N_\p(r<a_\star)}
\label{eq:dplstar}
\ee
where $\beta_v=1.8$ and $\frac{|\delta_\p \mathbf{L}|}{L_c}\simeq(1-e_\star^2)^{1/2}(\delta_\p  
I_\star^2+\sin I_\star^2\delta_\p\Omega_\star^2)^{1/2}$ is the relative variation of the angular 
momentum. After three orbits of the target star, simulations of the cluster M1 (or M2) result in
$\frac{|\delta_\p \mathbf{L}|}{L_c}\simeq 0.11'$ (or $0.04'$). These results are consistent with  
the predictions of Equation~\ref{eq:dplstar},  which are $0.1'$ and $0.03'$ for cluster M1 and M2, 
respectively.

Similar to Section \ref{subsubsec:pb_orb_S2}, we found that in most of the performed MC 
simulations, the Newtonian perturbations on the observables of the target star are 
mainly caused by the perturbed orbital period, rather than the Newtonian orbital precessions. 
Take the target star T1 as an example. From Table~\ref{tab:t5} we can see that $\langle 
\overline{|\delta_\p f_\star|}\rangle\gg\langle \overline{|\delta_\p I_\star|}\rangle$, $\langle 
\overline{|\delta_\p \Omega_\star|}\rangle$, or $\langle \overline{|\delta_\p 
\omega_\star|}\rangle$. By the rough estimations according to 
Equation~\ref{eq:r} and~\ref{eq:z}, ~\ref{eq:ar} and~\ref{eq:az}, it is simple to show that the 
perturbations on observables due to $\langle \overline{|\delta_\p f_\star|}\rangle$ are orders 
of magnitude larger than those due to $\langle\overline{|\delta_\p 
I_\star|}\rangle$, $\langle\overline{|\delta_\p \Omega_\star|}\rangle$ and 
$\langle\overline{|\delta_\p \omega_\star|}\rangle$.

\subsection{The critical semimajor axis}
~\label{subsec:cluster_crit}
From Figure~\ref{fig:sg_cluster_rms} and~\ref{fig:sg_cluster_zms}, we can see that the total 
perturbation is dominated by the spin-induced effects in the inner region of the cluster and 
the stellar perturbations in the outer region. For each model there is a critical orbital 
semimajor axis of the target star where $\langle \overline{|\delta_\p 
\mathbf{R}|}\rangle=\overline{|\delta_\s \mathbf{R}|}$ or  
$\langle \overline{\delta_\p Z}\rangle=\overline{\delta_\s Z}$. Denote $a_{\star}^{\rc}$ or 
$a_{\star}^{\zc}$ as the critical orbital semimajor axis for the position and redshift signals, 
respectively, then these values of the target star T1 or T2 and the corresponding pericenter distance 
($r_{{\rm per},\star}^{\rc}=a_{\star}^{\rc}(1-e_\star)$ and $r_{{\rm per},\star}^{\zc}=a_{\star}^{\zc}(1-e_\star)$, respectively)
in different models are showed in Table~\ref{tab:t4}. If target stars are with $a_\star$ larger than the critical 
value, the detection of the spin-induced effects from the observables is likely not feasible as 
they are quite significantly submerged by the stellar perturbations. 

As both the Newtonian and spin-induced perturbations are less significant for target stars 
with low orbital eccentricities, the resulting critical orbital semimajor axis of the target star 
T1 is found similar to those of the target star T2. For the clusters explored here, 
$a_{\star}^{\rc}\simeq120-390\AU$ and $a_{\star}^{\zc}\simeq100-330\AU$ for 
the star T1 or $a_{\star}^{\rc}\simeq100-300\AU$ and $a_{\star}^{\zc}\simeq110-350\AU$ for
the star T2. As the different MC realizations lead to about $\sim1$\,dex scatters on the stellar 
perturbations, the critical orbital semimajor axis given above may vary up to $25\%$. 

Note that the critical orbital semimajor axis depends also on the spin parameters (its magnitude and 
orientation). The spin orientation assumed here ($i=45^\circ$, $\epsilon=180^\circ$) exhibit 
modest spin-induced effects. If alternatively choose some other spin magnitude or orientations, 
then the orbital semimajor axis will be changed accordingly. Take target star T1 as an example. If 
the spin magnitude of the MBH is smaller, i.e., $a=0.3$, then for the position signal (or redshift 
signal), the critical semimajor axis in model M1-M6 are given by 
$a_{\star}^{\rc}\simeq100\AU-330\AU$ (or $a_{\star}^{\zc} \simeq 60\AU-210\AU$); If we assume 
$a=0.99$, $i=72^\circ$ and $\epsilon=91^\circ$, such that the spin-induced position displacement is 
most significant, then $a_\star^{\rc}\sim200\AU-600\AU$; If we assume $a=0.99$, 
$i=38^\circ$ and $\epsilon=46^\circ$, such that the spin-induced redshift difference is most 
significant, then $a_\star^{\zc}\simeq 110\AU-340\AU$.

\section{Discussion}
~\label{sec:discussion}

The high order relativistic effects, e.g., frame-dragging and the quadrupole effects, can be used 
in verifying the theory of general relativity, testing the quasi-Kerr metrics and also the no-hair 
theory~\citep[e.g.,][]{Johannsen16}. However, they should be measured from orbital motion of a 
target star after a clean removal of the competing perturbations caused by other perturbers. The 
contaminations of the stellar perturbations cause the deviations of the motion of the target 
star from the GR predictions. If these contaminating signals are not well removed, they may cause 
systematic bias of the measured spin parameters or other parameters, e.g., the mass of the black 
hole and the distance to the GC, obtained by the Markov Chain Monte Carlo fitting scheme. We defer 
a complete simulation of a data reduction and covariance analysis to future studies.

The complexity of the separation between the stellar perturbation and GR spin effects resides in 
the fact that the orbits of the perturbers are unknown to the observer. It seems quite challenge 
to fit the observational data from a number of target stars to recover simultaneously the orbital 
parameters of all the surrounding perturbers. In a simple case 
that the gravitational perturbations can be regarded as a result of a smoothed and static 
potential, the effects of the stellar perturbations can then be well modeled and 
removed~\citep{Merritt10}, although it is not clear whether it is a good assumption as the 
number of stars/ stellar remnants in the very vicinity of the MBH are likely to be small. 
A wavelet analysis may also help to disentangle the spin effects from the stellar 
perturbations~\citep{Angelil14}.

Other dynamical processes can also cause similar shifts in the apparent position and 
redshift of the target star. For example, gravitational waves and tidal dissipation. 
These effects are negligible for the S-stars considered in this paper, as they are only
important for those S-stars in highly eccentric orbits and/or in extremely tight orbits, 
e.g., $a_{\star}\la10^3r_\g\simeq40\AU$~\citep{Psaltis12,Psaltis13,Psaltis15}. It is still 
possible that there is an IMBH in the GC, with a mass around $10^2-10^4\msun$. As the stellar 
perturbations approximately proportional to the mass of the IMBH (See 
Figure~\ref{fig:mass_S0-102}), it can create perturbations on the target star that completely drown 
the spin effects if its mass is large enough.

Dynamical simulations suggest that pulsars possibly exist in the intermediate vicinity 
around the MBH in the GC~\citep[e.g.,][]{Zhang14}. If they are close 
enough to the MBH, the high order GR effects, including the frame-dragging and quadrupole effects, 
can be probed by their precise timing signals. Detections of such
pulsars and the measurements of their signal can possibly be realized by the Square 
Kilometer Array (SKA)~\citep[e.g.][]{Pfahl04,Liu12,Psaltis16}. However, the problems of the stellar 
perturbations and also the removal of them in the timing signals should also be 
considered in these studies. 

On the other hand, the measurements of the stellar perturbations constrain simultaneously 
the mass profiles in the vicinity of the MBH in the GC, which is essential for
studies of various important dynamical problems in the GC, e.g., the formation and 
dynamical evolution histories of the S-stars and stellar 
remnants~\citep[e.g.][]{Merritt11,Perets09,Madigan09,Madigan11,Zhang13}, the tidal 
disruptions of the stars by the black holes~\citep[e.g.][]{Bromley12,Madigan11} and the rates of 
gravitational wave in-spirals~\citep[e.g.][]{Merritt11,Hopman06}. 

\section{Conclusion}
~\label{sec:conclusion}
The S-stars discovered in the close vicinity of the MBH in the GC are 
anticipated to provide tight constraints on the MBH spin and metric by continuously monitoring  
their orbits. However, the gravitational attractions of other stars and stellar remnants in this 
region may deviate the orbit of a target S-star from GR predictions; thus, adequately modeling and 
removing them is quite essential. To understand comprehensively the stellar perturbations, here we 
consider both the spin-induced relativistic effects and the Newtonian perturbations felt by the 
target stars, and their resulting perturbations on the observables of the target star, i.e., the 
apparent position in the sky plane and the redshift. 

The relativistic numerical methods adopted here rely upon the framework of ZLY15. The 
gravitational attractions of the background stars/stellar remnants are considered by 
an additional perturbation term in the Hamiltonian equations of motion of the target star. 
The apparent orbital motion and the observed redshift of a target star are obtained by 
the ray-tracing techniques. We find that the simulated variations of the orbital elements 
of the target star due to the Newtonian perturbations resulting from the method adopted in this 
work are generally consistent with those from the post-Newtonian method with 2.0 order corrections.

The investigations of the gravitational perturbations by a single perturber can provide 
helpful hints in understanding the properties and the nature of the stellar 
perturbations. In this case, the Newtonian perturbations on S2/S0-2 caused by S0-102 are of 
particular interest. We find that the spin-induced effects on image position (or redshift) can 
be blurred by the gravitational perturbations from S0-102 alone if the mass of the latter is 
$\ga0.6\msun$ (or $\ga0.2\msun$), which is very likely as the S-stars are found exclusively B-type 
stars with masses $\ga3\msun$. 

The changes of the observables of the S2/S0-2 result from the 
changes of its orbital elements. We find that the perturbed observables of S2/S0-2, which is caused 
by the Newtonian attractions of S0-102, are mainly ascribed to the change of its orbital period. 
Meanwhile, the Newtonian orbital precessions due to S0-102 induce negligible difference on the 
observables of S2/S0-2. As a result, the Newtonian perturbations on the observables of S2/S0-2 peak 
around the time of the pericenter passage in each orbit and evolve quite differently with those of 
the spin-induced effects. We find that these conclusions also remain true for the general cases 
that a target star is perturbed by a single or multiple disturbing object(s).

By performing a large number of MC simulations, we study the case of a hypothetical 
S-star inside the orbits of S2/S0-2 or S0-102 perturbed by a single perturber with various 
initial conditions. We find that the Newtonian perturbations on the observables of the target star 
are proportional to the mass of the perturber, and depend complexly on the orbital configurations 
of both the perturber and the target star. The Newtonian perturbations of a single perturber 
located inside the orbit of the target star are found to be much more significant than those by a 
perturber located outside. It is found that in some cases the Newtonian perturbations on the 
observables due to a single perturber with mass of $10\msun$ is large enough to overwhelm the 
spin-induced effects (See Figure~\ref{fig:sg_a_T1} and~\ref{fig:sg_a_T2}).

So far the mass distribution and its composition in the vicinity of the MBH in 
the GC are rather uncertain. By performing a large number of numerical simulations that consider 
a number of possible initial conditions, we investigate the stellar perturbations of a cluster of 
disturbing objects. We find that, for a given total mass of the cluster, the stellar perturbations 
due to a cluster of stellar mass black holes (with masses of $10\msun$) are larger than those due 
to a cluster of low mass main-sequence stars (with masses of $1\msun$). The stellar 
perturbations of a cluster with the cusp profile are generally larger than those with the core 
profile, because it contains more stars in the inner region. When the central MBH is maximally 
spinning, the Newtonian perturbation of a cluster can drown the spin-induced signals if the target 
star is with orbital semimajor axis larger than $100-400\AU$. 

As showed in the numerical simulations performed in this study, the morphologies of the stellar 
perturbations seem quite different from the GR spin effects in both the evolutions of the 
orbital elements and the observables, i.e., the signals of the apparent position and the redshift.
Their different features and also the dependences on the model parameters suggest that, in 
principle, the stellar and the spin-induced perturbations are separable. We defer the 
separations of these two effects and the accurate measurements of the spin parameters in presence of 
the Newtonian perturbations to future studies. 

\acknowledgements
\noindent  We thank Lu Youjun, Yu Qingjuan, and Yang Yi-Jung for helpful suggestions.  This work 
was supported in part by the National Natural Science Foundation of China under grant Nos. 11603083, 11373031, the 
Fundamental Research Funds for the Central Universities grand No. 161GPY51. This work was also performed 
in part at the Aspen Center for Physics,  which is supported by National Science Foundation grant PHY-1066293. 
Part of the numerical works were performed in the computing cluster in School of Physics and Astronomy, Sun 
Yat-Sen University.

\end{document}